%% file: paper.tex
\documentclass[sigconf]{acmart}
\AtBeginDocument{%
  \providecommand\BibTeX{{%
    \normalfont B\kern-0.5em{\scshape i\kern-0.25em b}\kern-0.8em\TeX}}}

\copyrightyear{2022}
\acmYear{2022}
\setcopyright{rightsretained}

\acmConference[ISCA '22]{The 49th Annual International Symposium on Computer Architecture}{June 18--22, 2022}{New York, NY, USA}
\acmBooktitle{The 49th Annual International Symposium on Computer Architecture (ISCA '22), June 18--22, 2022, New York, NY, USA}
\acmPrice{}
\acmDOI{10.1145/3470496.3527415}
\acmISBN{978-1-4503-8610-4/22/06}

\usepackage{subfigure}

\usepackage{xspace}

\usepackage{hhline}
\usepackage{enumitem}
\usepackage{balance}
\usepackage{tabularx}
\usepackage{booktabs}
\usepackage{xfrac}
\usepackage{multirow}
\usepackage{threeparttable}
\usepackage{comment} 

\newcommand{\minitab}[2][l]{\begin{tabular}{#1}#2\end{tabular}}

\newcommand{\bhl}[1]{#1\xspace}

\newcommand{\name}{BTS\xspace}

\newcommand{\HAdd}{HAdd\xspace}
\newcommand{\HMult}{HMult\xspace}
\newcommand{\HRot}{HRot\xspace}
\newcommand{\HRes}{HRescale\xspace}
\newcommand{\PMult}{PMult\xspace}
\newcommand{\PAdd}{PAdd\xspace}
\newcommand{\CMult}{CMult\xspace}
\newcommand{\CAdd}{CAdd\xspace}
\newcommand{\evk}{$\mathbf{evk}$\xspace}
\newcommand{\evks}{$\mathbf{evk}$s\xspace}
\newcommand{\ct}{$\mathbf{ct}$\xspace}
\newcommand{\cts}{$\mathbf{ct}$s\xspace}
\newcommand{\scratchpad}{scratchpad\xspace}
\newcommand{\scratchpads}{scratchpads\xspace}
\newcommand{\dnum}{$\mathtt{dnum}$\xspace}
\newcommand{\dnums}{$\mathtt{dnum}$s\xspace}

\newcommand{\amormultslot}{T\textsubscript{mult,a/slot}\xspace}

\newcommand{\BrU}{BrU\xspace}

\newcommand{\npe}{n\textsubscript{PE}\xspace}
\newcommand{\npever}{n\textsubscript{PEver}\xspace}
\newcommand{\npehor}{n\textsubscript{PEhor}\xspace}

\newcommand{\nttz}{NTT\textsubscript{z}\xspace}
\newcommand{\ntty}{NTT\textsubscript{y}\xspace}
\newcommand{\nttx}{NTT\textsubscript{x}\xspace}
\newcommand{\lsub}{l\textsubscript{sub}\xspace}
\newcommand{\amtps}{$T_{mult,a/slot}$\xspace}

\newcommand{\lattigo}{{\small \texttt{Lattigo}}\xspace}
\newcommand{\hundredx}{{\small \texttt{100x}}\xspace}
\newcommand{\fone}{{\small \texttt{F1}}\xspace}
\newcommand{\foneplus}{{\bhl{\small \texttt{F1+}}}\xspace}
\newcommand{\nameone}{{\bhl{\small \texttt{INS-1}}}\xspace}
\newcommand{\nametwo}{{\bhl{\small \texttt{INS-2}}}\xspace}
\newcommand{\namethree}{{\bhl{\small \texttt{INS-3}}}\xspace}
\newcommand{\namex}{{\bhl{\small \texttt{INS-x}}}\xspace}

\usepackage{float}
\usepackage{aliascnt}
\newaliascnt{eqfloat}{equation}
\newfloat{eqfloat}{h}{eqflts}
\floatname{eqfloat}{Equation}

\newcommand*{\ORGeqfloat}{}
\let\ORGeqfloat\eqfloat
\def\eqfloat{%
  \let\ORIGINALcaption\caption
  \def\caption{%
    \addtocounter{equation}{-1}%
    \ORIGINALcaption
  }%
  \ORGeqfloat
}

\title{BTS: An Accelerator for Bootstrappable Fully Homomorphic Encryption}
\begin{document}

\author{Sangpyo Kim}
\email{vnb987@snu.ac.kr}
\affiliation{%
  \institution{Seoul National University}
  \city{Seoul}
  \country{South Korea}
}

\author{Jongmin Kim}
\email{jongmin.kim@snu.ac.kr}
\affiliation{%
  \institution{Seoul National University}
  \city{Seoul}
  \country{South Korea}
}

\author{Michael Jaemin Kim}
\email{michael604@snu.ac.kr}
\affiliation{%
  \institution{Seoul National University}
  \city{Seoul}
  \country{South Korea}
}

\author{Wonkyung Jung}
\email{wk@cryptolab.co.kr}
\affiliation{%
  \institution{Crypto Lab. Inc}
  \city{Seoul}
  \country{South Korea}
}

\author{John Kim}
\email{jjk12@kaist.edu}
\affiliation{%
  \institution{KAIST}
  \city{Daejeon}
  \country{South Korea}
}

\author{Minsoo Rhu}
\email{minsoo.rhu@gmail.com}
\affiliation{%
  \institution{KAIST}
  \city{Daejeon}
  \country{South Korea}
}

\author{Jung Ho Ahn}
\email{gajh@snu.ac.kr}
\affiliation{%
  \institution{Seoul National University}
  \city{Seoul}
  \country{South Korea}
}

\thispagestyle{plain}


\begin{abstract}
Homomorphic encryption (HE) enables the secure offloading of computations to the cloud by providing computation on encrypted data (ciphertexts).
HE is based on noisy encryption schemes in which noise accumulates as more computations are applied to the data.
The limited number of operations applicable to the data prevents practical applications from exploiting HE.
Bootstrapping enables an unlimited number of operations or \emph{fully} HE (FHE) by refreshing the ciphertext. 
Unfortunately, bootstrapping requires a significant amount of additional computation and memory bandwidth as well.
Prior works have proposed hardware accelerators for computation primitives of FHE.
However, to the best of our knowledge, this is the first to propose a hardware FHE accelerator that supports bootstrapping as a first-class citizen.

In particular, we propose \name\,---\,Bootstrappable, Technology-driven, Secure accelerator architecture for FHE.
We identify the challenges of supporting bootstrapping in the accelerator and analyze the off-chip memory bandwidth and computation required.
In particular, given the limitations of modern memory technology, we identify the HE parameter sets that are efficient for FHE acceleration.
Based on the insights gained from our analysis, we propose \name, which effectively exploits the parallelism innate in HE operations by arranging a massive number of processing elements in a grid.
We present the design and microarchitecture of \name, including a network-on-chip design that exploits a deterministic communication pattern.
\name shows 5,556$\times$ and 1,306$\times$ improved execution time on ResNet-20 and logistic regression over a CPU, with a chip area of 373.6mm\textsuperscript{2} and up to 163.2W of power.

\end{abstract}



\maketitle

\input{1_introduction}
\input{2_background}
\input{3_contribution-1}
\input{4_contribution-2}
\input{5_contribution-3}

\input{6_evaluation}

\input{7_relatedwork}
\input{8_conclusion}


\begin{acks}
This work was supported in part by Institute of Information \& communications Technology Planning \& Evaluation (IITP) grant funded by the Korea government (MSIT) (No. 2020-0-00840, 40\%) and National Research Foundation of Korea (NRF) grant funded by the Korean government (MSIT) (No. 2020R1A2C2010601, 60\%).
The EDA tool was supported by the IC Design Education Center (IDEC), Korea. 
Sangpyo Kim is with the Department of Intelligence and Information, Seoul National University.
Jung Ho Ahn, the corresponding author, is with the Department of Intelligence and Information, the Institute of Computer Technology, and the Research Institute for Convergence Science, Seoul National University, Seoul, South Korea.
\end{acks}

\bibliographystyle{ACM-Reference-Format.bst}
\bibliography{refs}

\end{document}

%% file: 1_introduction.tex
\section{Introduction}
\label{sec:1_introduction}

Homomorphic encryption (HE) allows computations on encrypted data or  ciphertexts (\cts).
In the machine-learning-as-a-service (MLaaS) era, HE is highlighted as an enabler for privacy-preserving cloud computing, as it allows safe offloading of private data.
Because HE schemes are based on the learning-with-errors (LWE)~\cite{jacm-2009-lwe} problem, they are noisy in nature.
Noise accumulates as we apply a sequence of computations on \cts.
This limits the number of computations that can be performed and hinders the applicability of HE for practical purposes, such as in deep-learning models with high accuracy~\cite{arxiv-2021-resnet20}.
To overcome this limitation, \emph{fully HE} (FHE)~\cite{stoc-2009-gentry-fhe} was proposed, featuring an operation (op) called \emph{bootstrapping}, that ``refreshes'' the \ct and hence permits an unlimited number of computations on the \ct.
Among multiple HE schemes that support FHE, CKKS~\cite{asia-2017-ckks} is one of the prime candidates as it supports fixed-point real number arithmetic.

One of the main barriers to adopting HE has been its high computational and memory overhead.
New schemes~\cite{toct-2014-bgv, iacr-2012-bfv2, siam-2014-bfv1, jc-2020-tfhe, asia-2017-ckks} and algorithmic optimizations~\cite{rsa-2020-better, eurocrypt-2021-efficient, tetc-2019-bfv} (using the residue number system~\cite{sac-2018-frns-ckks, sac-2016-behz}) have reduced this overhead and resulted in a 1,000,000$\times$ speedup~\cite{eurocrypt-2021-efficient} at least compared to its first HE implementation~\cite{eurocrypt-2019-implement-gentry}.
However, even with such efforts, HE ops experience tens of thousands of slowdowns compared to unencrypted ops~\cite{access-2021-demystify}.
Attempting to tackle this, prior works have sought hardware solutions to accelerate HE ops, including CPU extensions~\cite{access-2021-demystify, wahc-2021-hexl}, GPU~\cite{tches-2021-100x, access-2020-privft, tetc-2019-bfv, tches-2018-fv}, FPGA~\cite{asplos-2020-heax, hpca-2019-roy, fccm-2020-sunwoong-ntt, reconfig-2019-sunwoong-modmult}, and ASIC~\cite{micro-2021-f1}.

\setlength{\tabcolsep}{1.4pt}
\begin{table}[tb!]
\begin{threeparttable}
  \centering
  \caption{\label{tab:prior_Works} 
  Comparing prior HE acceleration works with \name
  }
  \begin{tabular}{c|cccccc}
  \toprule
   & \multirow{3}{*}{\minitab[c]{Plat\\-form}} & Target & \multirow{3}{*}{\minitab[c]{Boot\\-strap\\-pable}} & Refreshed &
   \multirow{3}{*}{\minitab[c]{Data\\parallel\\-ism\textsuperscript{\ddag}}} & FHE mult  \\
   &  & \ct len & & slots\textsuperscript{\dag} per & & thruput  \\
   &  &  ($N$) & & bootstrap & &($1/s$)  \\
  \midrule
  \texttt{Lattigo}~\cite{github-lattigo} & CPU &  $2^{16}$  & {\large $\circ$} & 32,768 & - & 6-10K \\
  \texttt{100x}~\cite{tches-2021-100x}& GPU &  $2^{17}$ & {\large $\circ$} & 65,536 & SIMT  & 0.1-1M \\
  \cite{hpca-2019-roy} & FPGA &  $2^{12}$  & $\times$ & -  & rPLP & $\times$ \\
  \texttt{HEAX}~\cite{asplos-2020-heax} & FPGA &  $2^{14}$ & $\times$ & - & rPLP & $\times$ \\
  \texttt{F1}~\cite{micro-2021-f1} & ASIC &  $2^{14}$  & $\;\;\triangle^{*}$ & 1 & rPLP & 4K\\
  \texttt{BTS} & ASIC & $2^{17}$ & {\large $\circ$} & 65,536  & CLP & 20M \\
  \bottomrule
  \end{tabular}
  \begin{tablenotes}\small
  \item[\dag] Data elements that can be packed in a \ct for SIMD execution.
  \item[\ddag] Residue-polynomial-level parallelism (rPLP) and coefficient-level parallelism (CLP) can be exploited in parallelizing HE ops (Section~\ref{sec:sub_4_parallelism_HE_function}).
  \item[*] \texttt{F1} only supports single-slot bootstrapping which has low throughput. 
  \end{tablenotes}
  \vspace{-0.2in}
\end{threeparttable}
\end{table}
\setlength{\tabcolsep}{6pt}

However, prior acceleration works mostly targeted small problem sizes, with a small target N (the length of a \ct), and they are lacking in bootstrapping support. Bootstrapping, which is necessary to reduce the impact of noise, occurs frequently in most FHE applications and represents the highest expense. For example, bootstrapping occurs more than 1,000 times for a single ResNet-20 inference~\cite{arxiv-2021-resnet20} and each instance of bootstrapping can take dozens of seconds on the state-of-the-art CPU~\cite{github-lattigo} and hundreds of milliseconds on a GPU~\cite{tches-2021-100x}. Most prior custom hardware acceleration works~\cite{hpca-2019-roy, asplos-2020-heax} do not support bootstrapping at all, while \fone~\cite{micro-2021-f1} demonstrated a bootstrapping time for CKKS but with limited throughput (Table~\ref{tab:prior_Works}).

We propose \name, a bootstrapping-oriented FHE accelerator that is \textbf{B}ootstrappable, \textbf{T}echnology-driven, and \textbf{S}ecure.
First, we identify the limitations that are imposed by contemporary fabrication technology when designing an HE accelerator, analyzing the implications of various conflicting requirements for the performance and security of FHE under such a constrained design space.
This allows us to pinpoint appropriate optimization targets and requirements when designing the FHE accelerator.
Second, we build a balanced architecture on top of those observations; we analyze the characteristics of HE functions to determine the appropriate number of processing elements (PEs) and proper data mapping that balances computation and data movement when using our FHE-optimized parameters.
We also choose to exploit coefficient-level parallelism (CLP), instead of residue-polynomial-level parallelism (rPLP), to evade the load imbalance issue.
Finally, we devise a novel PE microarchitecture that efficiently handles HE functions including base conversion, and a time-multiplexed NoC structure that manages both number theoretic transform and automorphism functions.

Through these detailed studies, \name achieves a 5,714$\times$ speedup in multiplicative throughput against \fone, the state-of-the-art ASIC implementation, when bootstrapping is properly considered.
Also, \name significantly reduces the training time of logistic regression~\cite{aaai-2019-helr} compared to the CPU (by 1,306$\times$) and GPU (by 27$\times$) implementations, and can execute a ResNet-20 inference 5,556$\times$ faster than the prior CPU implementation~\cite{arxiv-2021-resnet20}.

In this paper, we make the following key contributions:
\setlist{nolistsep}
\begin{itemize}[noitemsep,leftmargin=0.135in]
  \item We provide a detailed analysis of the interplay of HE parameters impacting the performance of FHE accelerators.
  \item We propose \name, a novel accelerator architecture equipped with massively parallel compute units and NoCs tailored to the mathematical traits of FHE ops.
  \item \name is the first accelerator targeting practical bootstrapping, enabling unbounded multiplicative depth, which is essential for complex workloads.
\end{itemize}

%% file: 2_background.tex
\section{Background}
\label{sec:2_background}

We provide a brief overview of HE and CKKS~\cite{asia-2017-ckks} in particular. Table~\ref{tab:symbol} summarizes the key parameters and notations we use in this paper.

\begin{table}[tb!]
\centering
\caption{\label{tab:symbol} List of symbols used to describe CKKS~\cite{asia-2017-ckks}.}
\renewcommand{\arraystretch}{1.1} 
\begin{tabular}{ll}
\toprule
Symbol                            & Definition                                              \\
\midrule
$Q$                               & (Prime) moduli product $=\prod_{i=0}^{L} q_i$           \\
$q_0, ..., q_L$                   & (Prime) moduli                                          \\
$Q_0, ..., Q_{\mathtt{dnum} - 1}$ & Modulus factors                                         \\
$P$                               & Special (prime) moduli product $=\prod_{i=0}^{k-1} p_i$ \\
$p_0, ..., p_{k - 1} $            & Special (prime) moduli                                  \\
$\mathbf{evk}_{mult}$             & Evaluation key ($\mathbf{evk}$) for HMult               \\
$\mathbf{evk}_{rot}^{(r)}$        & $\mathbf{evk}$ for HRot with a rotation amount of $r$   \\
$N$                               & The degree of a polynomial                              \\
$L$                               & Maximum (multiplicative) level                          \\
$\ell$                            & Current (multiplicative) level of a ciphertext          \\
$L_{boot}$                        & Levels consumed at bootstrapping                        \\
$k$                               & The number of special prime moduli                      \\
$\mathtt{dnum}$                   & Decomposition number                                    \\
$\lambda$                         & Security parameter of a given CKKS instance             \\
\bottomrule
\end{tabular}
\vspace{-0.05in}
\end{table}

\subsection{Homomorphic Encryption (HE)}
\label{sec:sub_2_homomorphic_encryption}

HE enables direct computation on encrypted data, referred to as ciphertext (\ct), without decryption.
There are two types of HE.
\emph{Leveled HE (LHE)} supports  a limited number of operations (ops) on a \ct due to the noise that accumulates after the ops.
In contrast, \emph{Fully HE (FHE)} allows an unlimited number of ops on \cts through bootstrapping~\cite{stoc-2009-gentry-fhe} that ``refreshes'' a \ct and lowers the impact of noise.
LHE has limited applicability\footnote{The hybrid use of LHE with multi-party computation~\cite{crypto-2012-mpc} allows for a broader range of applications. However, such an approach has a different bottleneck of the communication cost and intense client-side computations.}; in the field of privacy-preserving deep learning inference, for instance, simple/shallow networks such as LoLa~\cite{pmlr-2019-lola} can be implemented with LHE, but only with limited accuracy (74.1\%). 
More accurate models such as ResNet-20~\cite{arxiv-2021-resnet20} (92.43\%) demand much more ops applied to \cts and thus FHE implementation.

While other FHE schemes support integer~\cite{toct-2014-bgv, siam-2014-bfv1, iacr-2012-bfv2} or boolean~\cite{jc-2020-tfhe} data types,
CKKS~\cite{asia-2017-ckks} supports fixed-point complex (real) numbers.
As many real-world applications such as MLaaS (Machine Learning as a Service) require arithmetic on real numbers, CKKS has become one of the most prominent FHE schemes.
In this paper, we focus on accelerating CKKS ops;
however, our proposed architecture is applicable to other popular FHE schemes (e.g., BGV~\cite{toct-2014-bgv} and 
BFV~\cite{siam-2014-bfv1,iacr-2012-bfv2, sac-2016-behz}) that share similar core ops.

\subsection{CKKS: an emerging HE scheme}

\label{sec:sub_2_CKKS_overview}

CKKS first encodes a message that is a vector of complex numbers, into a plaintext $ m(X)\! =\! \sum_{i=0}^{N-1} c_i X^i $, which is a polynomial in a cyclotomic polynomial ring $ \mathcal{R}_Q\! =\! \mathbb{Z}_Q[X]/(X^N\! + 1) $.
The coefficients $ \{c_i\} $ are integers modulo $ Q $ and the number of coefficients (or degree) is $ N $,
where $ N $ is a power-of-two integer, typically ranging from $ 2^{10} $ to $ 2^{18} $.
For a given $ N $, a message with up to $ \sfrac{N}{2} $ complex numbers
can be \emph{packed}
into a single plaintext in CKKS.
Each element within a packed message is referred to as a \emph{slot}.
After encoding (or packing), element-wise multiplication (mult) and addition between two messages can be done through polynomial operations between plaintexts.  
CKKS then encrypts a plaintext $m(X) \in \mathcal{R}_Q$ into a $\mathbf{ct}\in \mathcal{R}^2_Q$ based on the following equation,
\begin{equation*}
\begin{split}
\mathbf{ct}\! =\! (b(X), a(X))\! =\! (a(X) \cdot s(X) + m(X) + e(X), a(X))
\end{split}
\end{equation*}
\vspace{-0.15in}

\noindent 
where  $s(X) \! \in \! \mathcal{R}_Q$ is a secret key, 
$a(X) \! \in \! \mathcal{R}_Q$ is a random polynomial, and 
$e(X)$ is a small Gaussian error polynomial required for LWE security guarantee~\cite{iacr-2019-standard}.
CKKS decrypts \ct by computing $ m^{\prime}(X)\! =\! \mathbf{ct} \boldsymbol{\cdot} (1, -s(X))\!  =\! m(X)\! +\! e(X) $, which approximates to $ m(X) $ with small errors.

HE is mainly bottlenecked by the high computational complexity of polynomial ops.
As each coefficient of a polynomial is a large integer (up to 1,000s of bits) and the degree is high (even surpassing 100,000), an op between two polynomials has high compute and data-transfer costs.
To reduce the computational complexity, HE schemes using the residue number system (RNS)~\cite{sac-2016-behz, sac-2018-frns-ckks} have been proposed.
For example, Full-RNS CKKS~\cite{sac-2018-frns-ckks} sets $ Q $ as the product of word-sized \emph{(prime) moduli} $\{q_i\}_{0 \leq i \leq L} $, where $ Q\! =\! \prod_{i=0}^{L} q_i $ for a given integer $ L $.
Using the Chinese remainder theorem (Eq.~\ref{eq:CRT}), we represent a polynomial in $ \mathcal{R}_Q $ with \emph{residue polynomials} in $ \{ \mathcal{R}_{q_i} \}_{0 \leq i \leq L} $, whose coefficients are residues obtained by performing modulo $ q_i $ (represented as $[\cdot]_{q_i}$) on the large coefficients:
\begin{equation}
\begin{split}
    [a(X)]_Q\! \mapsto\! ([a(X)]_{q_{0}}, \hdots, [a(X)]_{q_{L}}) \text{ where } Q\! =\! \prod_i q_i
\end{split}
\label{eq:CRT}
\end{equation}
\vspace{-0.15in}

\noindent Then, we can convert an op involving two polynomials into ops between the residue polynomials with word-sized coefficients ($\leq$ 64 bits) corresponding to the same $ q_i $, avoiding costly big-integer arithmetic with carry propagation.
Full-RNS CKKS provides an $\sim$8$\times$ speedup over plain CKKS~\cite{sac-2018-frns-ckks} and thus, we adopt Full-RNS CKKS as our CKKS implementation, representing a polynomial in  $ \mathcal{R}_Q $ as an $ N \! \times \! (L\! +\! 1) $ matrix of residues, and a \ct as a pair of such matrices.

\subsection{Primitive operations (ops) of CKKS}
\label{sec:sub_2_primary_CKK_operations}

Primitive HE ops of CKKS are introduced here, which can be combined to create more complex HE ops such as linear transformation and convolution.
Given two ciphertexts $ \mathbf{ct}_0,\mathbf{ct}_1$ where $\mathbf{ct}_i\!=\!(b_i(X),a_i(X))$ and $b_i(X)=a_i(X)\cdot s(X)+m_i(X)$, HE ops can be summarized as follows:

\textbf{\HAdd} performs an element-wise addition of $ \mathbf{ct}_0 $ and $ \mathbf{ct}_1 $: 
\begin{equation}
        \mathbf{ct}_{add} = (b_0(X) + b_1(X), a_0(X) + a_1(X)) 
\end{equation}
\vspace{-0.15in}

\textbf{\HMult} consists of a \emph{tensor product} and \emph{key-switching}.
The tensor product first creates $ (d_0(X), d_1(X) , d_2(X)) $:
\begin{equation}
    \begin{split}
       d_0(X) & = b_0(X) \cdot b_1(X) \\
       d_1(X) & = a_0(X) \cdot b_1(X) + a_1(X) \cdot b_0(X)\\
       d_2(X) & = a_0(X) \cdot a_1(X)
    \end{split}
\end{equation}
\vspace{-0.05in}

\noindent
By computing $(d_0(X), d_1(X), d_2(X)) \boldsymbol{\cdot} (1, -s(X), s(X)^2) $, we recover $ m_0(X) \cdot m_1(X) $, albeit with error terms.
Key-switching recombines the tensor product result to be decryptable with $(1, -s(X))$ using a public key, called an evaluation key (\evk).
An \evk is a \ct in $\mathcal{R}_{PQ}^2$ with a larger modulus $PQ$, where 
$P=(\prod_{i=0}^{k-1} p_i) \geq Q$ for given \emph{special (prime) moduli} $p_0, \ldots, p_{k-1}$.
We express an \evk as a pair of $ N\! \times\! (k\! +\! L\! +\! 1) $ matrices.
\HMult is then computed using Eq.~\ref{eq:ctmult}, which involves key-switching with an $\mathbf{evk}$ for mult, $\mathbf{evk}_{mult}$:
\begin{equation}
\mathbf{ct}_{mult} = (d_0(X), d_1(X))+\underbrace{P^{-1} (d_2(X) \cdot \mathbf{evk}_{mult})}_\text{key-switching}
\label{eq:ctmult}
\end{equation}

\textbf{\HRot} circularly shifts a message vector by slots.
When a $\mathbf{ct}$ encrypts a message vector $\mathbf{z}$\! =\! $(z_0, ... , z_{N/2 - 1})$, after applying HRot by a \emph{rotation amount} $ r $, the rotated ciphertext $\mathbf{ct}_{rot}$ encrypts $\mathbf{z}^{(r)}$\! =\! $(z_r, ..., z_{N/2 - 1}, z_{0}, ..., z_{r-1})$.
HRot consists of an \emph{automorphism} and key-switching.
$\mathbf{ct}\! =\! (b(X), a(X))$ is mapped to $\mathbf{ct}^{\prime}\! =\! (b(X^{5^r}), a(X^{5^r}))$ after an automorphism.
This moves the coefficients of a polynomial through the mapping ${i \mapsto \sigma_r(i)}$, where $i$ is the index of the coefficient $c_i$ and $\sigma_r$ is:
\begin{equation}
\sigma_r : i \mapsto i \cdot 5^r \text{ mod } N\ \ (i = 0, 1, \ldots, N - 1 )
\label{eq:automorphism}
\end{equation}
\vspace{-0.15in}

Similar to \HMult, key-switching brings back $\mathbf{ct}^{\prime}$, which was only decryptable with $(1, -s(X^{5^r}))$ after
automorphism, to be decryptable with $(1, -s(X))$.
An \HRot with a different rotation amount each requires a separate $\mathbf{evk}$, $\mathbf{evk}_{rot}^{(r)}$.
\HRot is computed as follows:
\begin{equation}
\mathbf{ct}_{rot} =(b(X^{5^r}), 0) + P^{-1} ( a(X^{5^r}) \cdot \mathbf{evk}_{rot}^{(r)})
\end{equation}
\vspace{-0.15in}

HE applications require other HE ops, such as an addition or mult of a \ct
with a scalar (\textbf{\CAdd}, \textbf{\CMult}) or a polynomial (\textbf{\PAdd}, \textbf{\PMult}) of unencrypted, constant values.
Additions are performed by adding the scalar or polynomial to $b(X)$, and mults are performed by multiplying each $b(X)$ and $a(X)$ by the scalar or polynomial.

\subsection{Multiplicative level and HE bootstrapping}
\label{sec:sub_2_mult_level_bootstrapping}

\noindent
\textbf{Multiplicative level}:
The error included in a \ct is amplified during HE ops;
in particular, \HMult multiplies the error $e(X)$ with other terms (e.g., $ m_0(X)$ and $m_1(X)$) and can result in an explosion of the error if not treated properly. 
CKKS performs \textbf{\HRes} to mitigate this explosion and keep the error tolerable by dividing the \ct by the last prime modulus $q_L$~\cite{sac-2018-frns-ckks}.
After \HRes, the  $q_L$  residue polynomial is discarded, and the \ct is reduced in size.
The \ct continues losing  the residues of $q_{L - 1}, \ldots, q_1$ with each 
\HRes
while executing an HE application   
until only one residue polynomial 
is left 
when no additional \HMult can be performed on the \ct. 
$L$, or the \emph{maximum multiplicative level}, determines the maximum number of \HMult ops that can be performed without bootstrapping, and the current \emph{(multiplicative) level} $\ell$ denotes the number of remaining \HMult operations that can be performed on  the \ct. 
Thus, a \ct with a level $\ell$ is represented as a pair of $ N\! \times\! (\ell\! +\! 1) $ matrices.

\noindent
\textbf{Bootstrapping}:
FHE features a \emph{bootstrapping} op that restores the multiplicative level ($\ell$) of a \ct to enable more ops.
Bootstrapping must be commonly performed for the practical usage of HE with a complex sequence of HE ops.
Bootstrapping mainly consists of homomorphic linear transforms and approximate sine evaluation~\cite{eurocrypt-2018-heaanboot}, which can be broken down into hundreds of primitive HE ops.
HMult and HRot ops account for more than 77\% of the bootstrapping time~\cite{github-lattigo}.
As bootstrapping itself consumes $L_{boot}$ levels, $L$ should be larger than $L_{boot}$. A larger $L$ is beneficial as it requires less frequent bootstrapping.
$L_{boot}$ ranges from 10 to 20 depending on the bootstrapping algorithm; a larger $L_{boot}$ allows the use of more precise and faster bootstrapping algorithms\cite{eurocrypt-2019-improved,eurocrypt-2021-efficient, eurocrypt-2021-invsine, rsa-2020-better}.
The bootstrapping algorithm we use in this paper is based on \cite{rsa-2020-better} with updates to meet the latest security and precision requirements~\cite{eurocrypt-2021-efficient, iacr-2020-varmin, access-2019-hybridattack}, and the value of $L_{boot}$ is 19.
Readers are encouraged to refer to the papers for a more detailed explanation of the algorithm.
Another CKKS-specific constraint is that the moduli $q_i$'s and the special moduli $p_i$'s must be large enough to tolerate the error accumulated during bootstrapping, whose typical values range from $2^{40}$ to $2^{60}$~\cite{iacr-2021-cheon-practical, github-lattigo}.

\subsection{Modern algorithmic optimizations in CKKS and amortized mult time per slot}
\label{sec:sub_2_modern_algo_CKKS}

\noindent
\textbf{Security level ($\lambda$)}:
The level of security for an HE scheme is represented by $\lambda$, a parameter measured by the logarithmic time complexity for an
attack~\cite{access-2019-hybridattack} to deduce the secret key.
A sufficiently high $\lambda$ is required for safety;
we target $\lambda$ of 128 bits, adhering to the standard~\cite{iacr-2019-standard} established by recent HE studies~\cite{eurocrypt-2021-efficient, eurocrypt-2021-invsine, iacr-2020-varmin} and libraries~\cite{github-lattigo, online-palisade}.
$\lambda$ is a strictly increasing function of $\sfrac{N}{\log PQ}$~\cite{wahc-2019-curtis}.

\noindent
\textbf{Dnum}:
Key-switching is an expensive function, accounting for most of the time in \HRot and \HMult~\cite{tches-2021-100x}.
We adopt a state-of-the-art generalized key-switching technique~\cite{rsa-2020-better}, which balances $L$, the computational cost, and $\lambda$.
\cite{rsa-2020-better} factorizes the moduli product $Q$ into $Q\! =\! Q_0 \cdot ... \cdot Q_{\mathtt{dnum}-1}$ (see Eq.~\ref{eq:q_factorization}) for a given integer \dnum (\emph{decomposition number}). It decomposes a \ct into \dnum slices, each consisting of residue polynomials corresponding to the prime moduli ($q_i$'s) that together compose the modulus factor $Q_j$.
We perform key-switching on each slice in $\mathcal{R}_{Q_j}$ and later accumulate them.
The special moduli product $P$ should only satisfy $P \geq Q_j$ for each $Q_j$, allowing us to choose a smaller $P$, leading to a higher $\lambda$.
i) Therefore, a larger \dnum means \emph{a greater level} of $L$ with fixed values of $\lambda$ and $N$ because we can increase $Q$.
\begin{equation}
    \begin{split}
        Q = \underbrace{q_0\! \cdot\! ...\! \cdot\! q_{\frac{L\! +\! 1}{\mathtt{dnum}}\! -\! 1}}_{Q_0}\! \cdot\! \underbrace{q_{\frac{L\! +\! 1}{\mathtt{dnum}}}\! \cdot\! ...\! \cdot\! q_{2 \frac{L\! +\! 1}{\mathtt{dnum}}\! -\! 1}}_{Q_1}\! \cdot ...\! \cdot\! \underbrace{q_{(\mathtt{dnum}\! -\! 1) \frac{L\! +\! 1}{\mathtt{dnum}} }\! \cdot\! ...\! \cdot\! q_{L\! +\! 1}}_{Q_{\mathtt{dnum\!-\!1}}}
    \end{split}
\label{eq:q_factorization}
\end{equation}
\vspace{-0.07in}

A major challenge of generalized key-switching is that different $\mathbf{evk}$s ($\mathbf{evk}_0, ..., \mathbf{evk}_{\mathtt{dnum} - 1}$) must be prepared for each factor $Q_j$, where each \evk is a pair of $N\! \times\! (k\! + \!L\! +\! 1) $ matrices and $k$ is set to $\sfrac{(L\! +\! 1)}{\mathtt{dnum}}$.
ii) Thus, the aggregate \evk size becomes $2 \cdot N \cdot (L \!+ \!1) \cdot (dnum \!+ \!1)$, linearly increasing with \dnum.
iii) The overall computational complexity of a single HE op also increases with \dnum.
Therefore, choosing an appropriate $\mathtt{dnum}$ crucially affects the performance.

\noindent
\textbf{Amortized mult time per slot (\amormultslot)}:
Changing the HE parameter set has mixed effects on the performance of HE ops.
Decreasing $N$ reduces the computational complexity and memory usage.
However, we should lower $L$ and $Q$ to sustain security, which requires more frequent bootstrapping.
Also, because a \ct of degree $N$ can encode only up to $N/2$ message \emph{slots} by packing, the throughput degrades.

Jung et al.\cite{tches-2021-100x} introduced a metric called \emph{amortized mult time per slot} (\amormultslot), which is calculated as follows:
\begin{equation}
\text{\amormultslot} = \frac{\text{T\textsubscript{boot}} + \sum_{\ell = 1}^{L - L_{boot}} \text{T\textsubscript{mult}}(\ell)}{L - L_{boot}} \cdot \frac{2}{N}
\end{equation}
\vspace{-0.07in}

\noindent where T\textsubscript{boot} is the bootstrapping time and T\textsubscript{mult}$(\ell)$ is the time required to perform HMult at a level $\ell$.
This metric initially calculates the average cost of mult including the overhead of bootstrapping, and then divides it by the number of slots in a \ct ($\sfrac{N}{2}$).
Thus, \amormultslot effectively captures the reciprocal throughput of a \emph{CKKS instance} (CKKS scheme with a certain parameter set).

%% file: 3_contribution-1.tex
\section{Technology-driven Parameter Selection of Bootstrappable Accelerators}
\label{sec:3_contribution1}

\subsection{Technology trends regarding memory hierarchy}
\label{sec:sub_3_tech_trend}

Domain-specific architectures (e.g., deep-learning~\cite{isca-2021-tpuv4i, micro-2020-gaudi, hcs-2021-graphcore} and multimedia~\cite{asplos-2021-youtube} accelerators) are often based on custom logic and an optimized dataflow to provide high computation capabilities.  
In addition, the memory capacity/bandwidth requirements of the applications are exploited in the design of the memory hierarchy. 
Recently, on-chip SRAM capacities have scaled significantly~\cite{iedm-2017-intel10nm} such that the level of hundreds of MBs of on-chip SRAM is feasible, providing tens of TB/s of SRAM bandwidth\cite{isca-2021-tpuv4i, hcs-2021-sambanova, hcs-2021-graphcore}. 
While the bandwidth of the main-memory has also increased, its aggregate throughput is still more than an order of magnitude lower than the on-chip SRAM bandwidth~\cite{micro-2017-finegrainedDRAM}, achieving a few TB/s of throughput even with high-bandwidth memory (HBM).

Similar to other domain-specific  architectures~\cite{isca-2016-eyeriss,isca-2021-tpuv4i}, HE applications also follow deterministic computational flows, and the locality of the input and output \cts of HE ops can be maximized through software scheduling~\cite{pldi-2020-eva}.  Thus, \cts can be reused by exploiting a large amount of on-chip SRAM enabled by technology scaling.  However, even with the increasing on-chip SRAM capacity, we observe that the size of on-chip SRAM is still insufficient to store \evks, rendering the off-chip memory bandwidth becomes a crucial bottleneck for modern CKKS scheme that supports bootstrapping.  In the following sections, we identify the importance of bootstrapping on the overall performance and provide an analysis of how different CKKS parameters impact the amount of data movement during bootstrapping and its final throughput.

\subsection{Interplay between key CKKS parameters}
\label{sec:sub_3_interplay}

\begin{figure}
\centering
  \subfigure[Maximum level $L$\label{fig:dnum_vs_level}] {\includegraphics[height=1.3in]{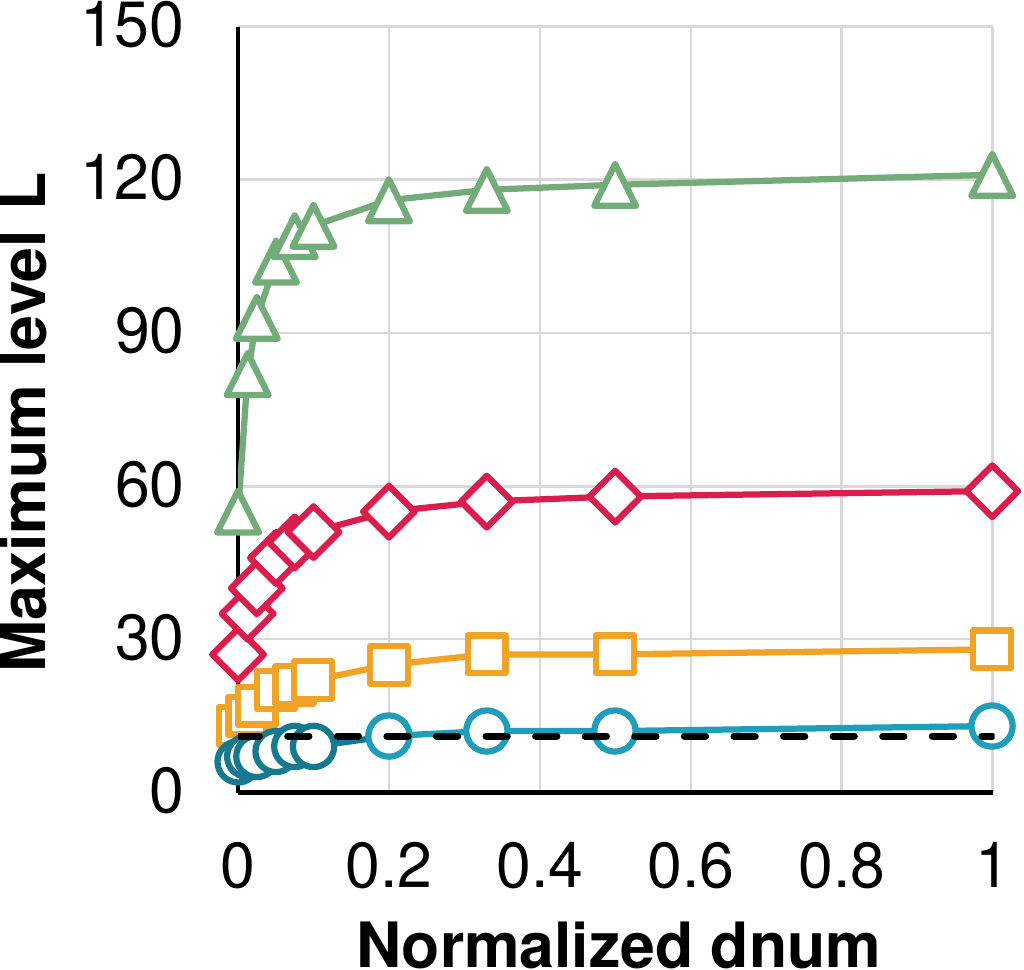}}
  \hfill
   \subfigure[$\mathbf{evk}$ size\label{fig:dnum_vs_evk}] {\includegraphics[height=1.3in]{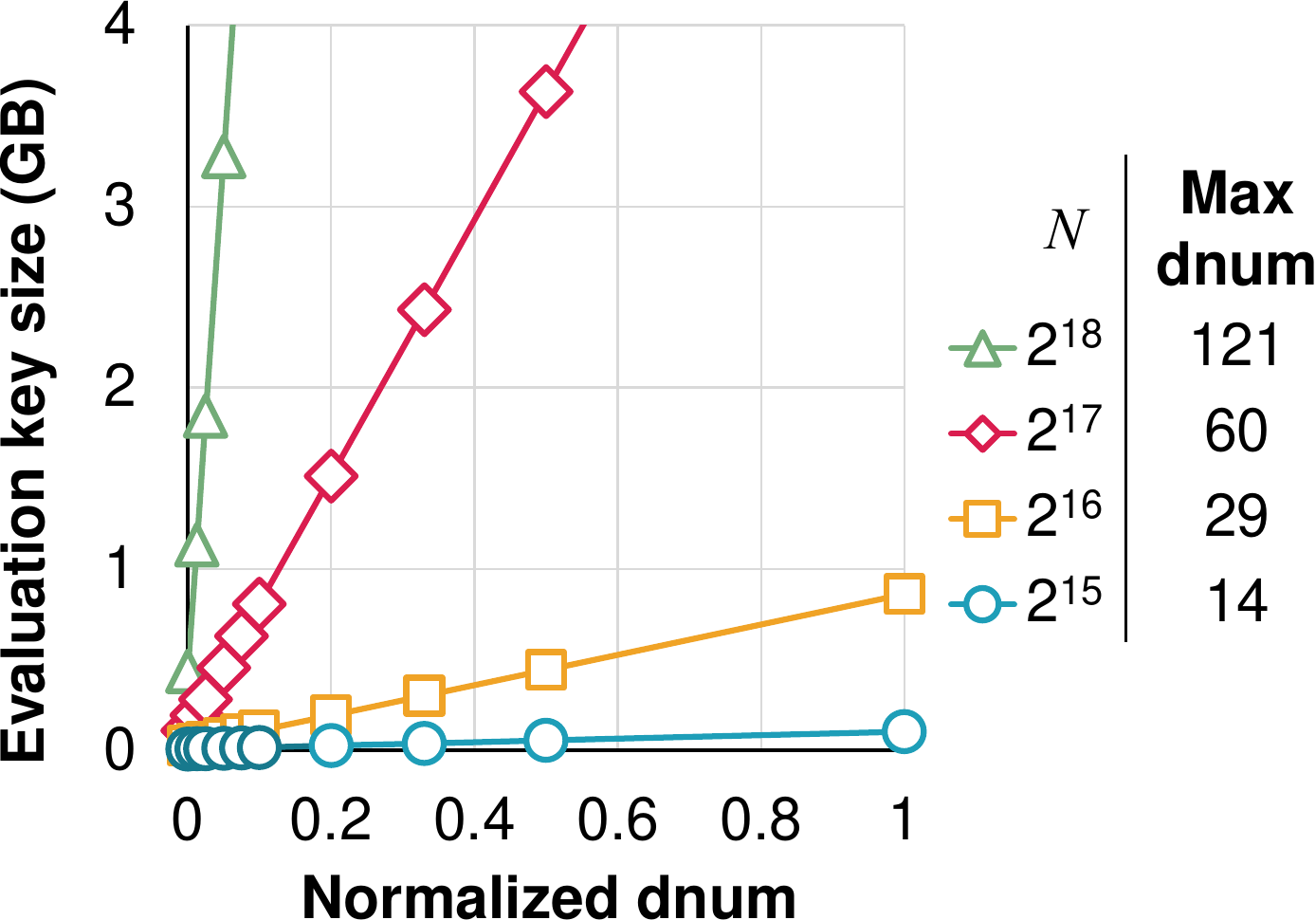}}
    \vspace{-0.1in}
    \caption{(a) $L$ and (b) a single \evk size vs. \dnum for four different $N$ (polynomial degree) values and a fixed 128b security target. 
    Normalized-\dnum of 0 means \dnum = 1 and normalized-\dnum of 1 means \dnum = max (i.e., $k$ = 1).
    Interpolated results are used for points with non-integer \dnum values. 
    The dotted line in (a) represents the minimum required level of 11 for bootstrapping.}
    \vspace{-0.12in}
    \label{fig:interplay_parameters}
\end{figure}

Selecting one parameter of a CKKS instance has a multifaceted effect on the other parameters. 
First, $\lambda$ is lowered when $Q$ is higher, and is raised when $N$ is higher. 
Considering that a bootstrappable CKKS instance requires a high $L$ ($>\! L_{boot}$), and with the sizes of prime moduli $q_i$ and $p_i$ set around $2^{50}$ and $2^{60}$ with a 64-bit machine word size, $\log PQ$ exceeds 500.
To support 128b security when $\log PQ$ exceeds 500, $N$ must be larger than $2^{14}$~\cite{iacr-2020-varmin}.

Second, when $\log PQ$ is set from fixed values of $\lambda$ and $N$, a larger \dnum leads to a higher $L$ at the cost of a larger \evk size.
Considering that $k$ equals $\sfrac{(L\! +\! 1)}{\mathtt{dnum}}$, the $Q\!:\!P$ ratio is close to $\mathtt{dnum}\!:\!1$.
Therefore, when $\log PQ$ is fixed, a larger \dnum means a larger $Q$ and finally a larger $L$. However, the \evk size also increases linearly with \dnum (see Fig.~\ref{fig:interplay_parameters}).
Because the high level of $L$ achieved by increasing \dnum saturates quickly, choosing a proper \dnum is important.

\begin{figure}[tb!]
    \begin{center}
    \includegraphics[width=0.99\columnwidth]{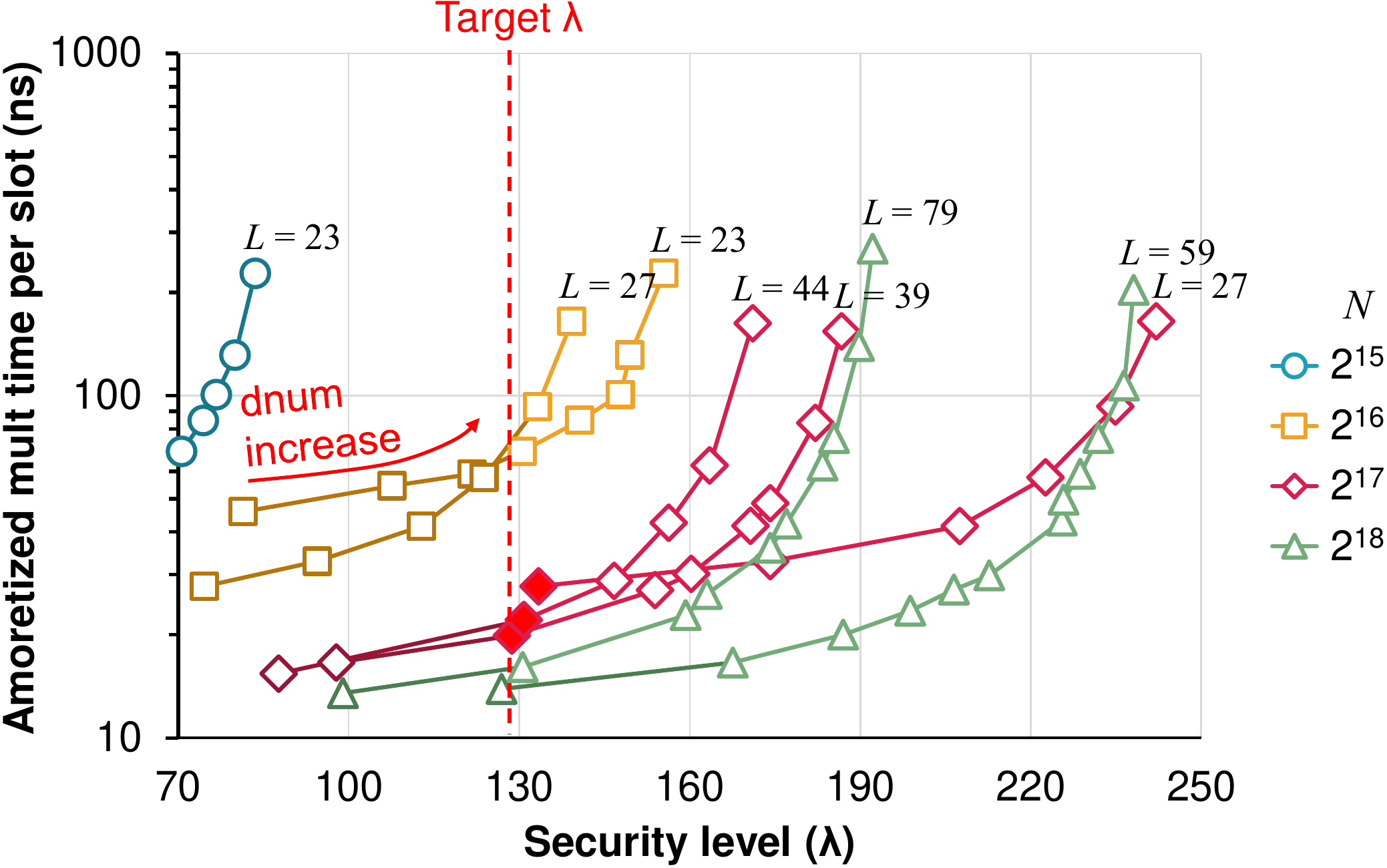}
    \end{center}
    \vspace{-0.05in}
    \caption{$\lambda$ and the minimum bound \amtps of an HE accelerator simulated for different CKKS instances.
    Results are measured for all possible integer \dnum values including 1 and the max for each (N, L) pair. The points highlighted in red represent ($N$, $L$, \dnum) = ($2^{17}$,~27,~1), ($2^{17}$,~39,~2), ($2^{17}$,~44,~3).}
    \label{fig:target_parameter}
    \vspace{-0.2in}
\end{figure}

\subsection{Realistic minimum bound of HE accelerator execution time}
\label{sec:sub_3_bootstrapping_analysis}

\amormultslot is mainly determined by the bootstrapping time, as bootstrapping is more than 60$\times$ longer than a single \HMult on conventional systems~\cite{github-lattigo, tches-2021-100x}.
Unlike simple LHE tasks such as LoLa~\cite{pmlr-2019-lola}, which only requires a handful of \evks, bootstrapping typically requires more than 40 \evks, mostly for the long sequence of multiple HRots applied with different $r$'s during the linear transformation steps of bootstrapping~\cite{eurocrypt-2021-efficient} ($\mathbf{evk}_{rot}^{(r)}$).
They can amount to GBs of storage and exhibit poor locality.

The bootstrapping time is mostly spent on HMult and HRot.
\cite{tches-2021-100x} found that \HMult and \HRot are memory-bound, highly dependent on the on-chip storage capacity.
Given today's technology with low logic costs and high-density on-chip SRAMs, the performance of \HMult and \HRot can be improved significantly with an HE accelerator.

Despite such an increase in on-chip storage, \evks, with each possibly taking up several hundreds of MBs (see Fig.~\ref{fig:interplay_parameters}), cannot easily be stored on-chip.
Because on-chip storage cannot hold all \evks,
they must be stored off-chip and be loaded in a streaming fashion upon every \HMult/\HRot.
Therefore, even if every temporal data and \cts with high locality are assumed to be stored on-chip with
massive on-chip storage,
the load time of \evk becomes the minimum execution time for \HMult/\HRot considering the limited off-chip bandwidth.

\begin{figure*}[tb!]
\centering
  \subfigure[Computational flow\label{fig:key_switch}] {\includegraphics[height=1.21in]{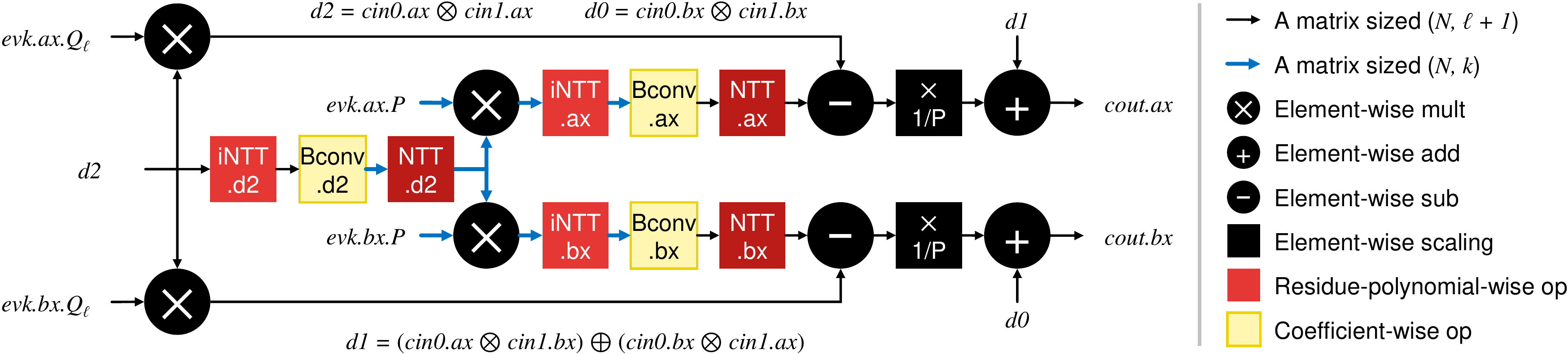}}
   \hfill
   \subfigure[Relative complexity\label{fig:complexity}] {\includegraphics[height=1.21in]{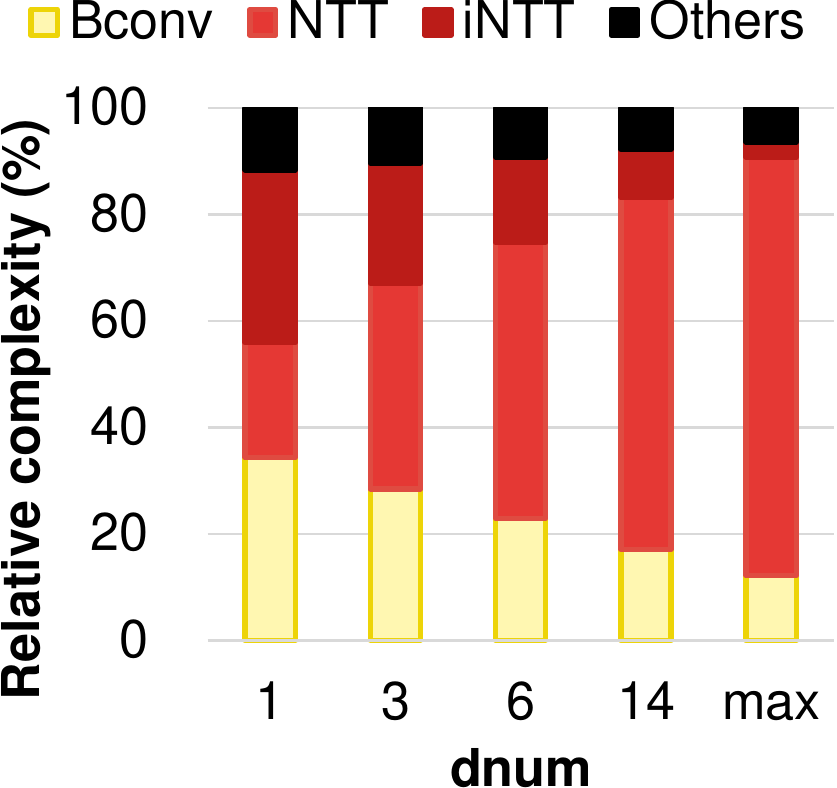}}
\vspace{-0.11in}
\caption{
  (a) Computational flow of the key-switching inside HMult and (b) computational complexity breakdown of HMult for \cts at the maximum level on CKKS instances with the same $N\!=\!2^{17}$ and $\lambda\!=\!128$ values but different \dnum values.
  The computational complexity is analyzed based on \cite{tches-2021-100x}.}\label{fig:key_switch_and_complexity}
\vspace{-0.05in}
\end{figure*}

\subsection{Desirable target CKKS parameters for HE accelerators}
\label{sec:sub_3_desirable_parameters}

To understand the impact of CKKS parameters, we simulate \amormultslot at multiple points while sweeping the $N$, $L$, and \dnum values. 
With 1TB/s of memory bandwidth (half of NVIDIA A100~\cite{micro-2021-a100} and identical to \fone~\cite{micro-2021-f1}), a bootstrapping algorithm that consumes 19 levels, and the simulation methodology 
in Section~\ref{sec:sub_6_experimental_setup}, we add two simplifying assumptions based on Section~\ref{sec:sub_3_bootstrapping_analysis} such that 1) the computation time of HE ops can be fully hidden by the memory latency of \evks, and 2) all \cts of HE ops are stored in on-chip SRAM and re-used.
Fig.~\ref{fig:target_parameter} reports the results.
The x-axis shows $\lambda$ determined by $\sfrac{N}{\log PQ}$~\cite{wahc-2019-curtis}, as calculated using an estimation tool~\cite{github-sparse-lwe-estimator}.
The y-axis shows \amormultslot for different $N$s, $L$s, and \dnums.

We make two key observations.
First, when other values are fixed, \amormultslot decreases as $N$ increases, even with the higher memory pressure from the larger \cts and \evks because the available level ($L\! -\! L_{boot}$) increases.
However, such an effect saturates after $N\! =\! 2^{17}$. 
Around our target security level of 128b in Fig.~\ref{fig:target_parameter}, the gain from $2^{16}$ to $2^{17}$ is 3.8$\times$ (111.4ns to 29.1ns), whereas that from $2^{17}$ to $2^{18}$ is 1.3$\times$.
Second, while a higher \dnum can help smaller $N$s to reach our target 128b security level, it comes at the cost of a superlinear increase in \amormultslot due to the increasing \evk size and the additional gain in $L$ being saturated.

These key observations suggest that a bootstrappable HE accelerator should target CKKS instances with \emph{high polynomial degrees} ($N\! \geq\! 2^{17}$) and \emph{low \dnum} values.
Our \name targets the CKKS instances with $N=2^{17}$ highlighted in Fig.~\ref{fig:target_parameter}.
With these,
the simulated HE accelerator achieves \amormultslot of 27.7ns, 19.9ns, and 22.1ns with corresponding ($L$,~\dnum) pairs of (27,~1), (39,~2), and (44,~3), respectively.
Although \name can support all CKKS instances shown in Fig.~\ref{fig:target_parameter}, it is not optimized for other CKKS instances as they either exhibit worse \amormultslot, or require significantly more on-chip resources with only a marginal performance gain ($N\!=\!2^{18}$).

In this paper, we use the CKKS instance with $N\! =\! 2^{17}$, $L\! =\! 27$, and $\mathtt{dnum}\! =\! 1$ as a running example.
When using the 64-bit machine word size, a $\mathbf{ct}$ at the maximum level has a size of 56MB, and an \evk has a size of 112MB.

%% file: 4_contribution-2.tex
\section{Architecting \name}
\label{sec:4_design_principles}

We explore the organization of \name, our HE accelerator architecture.
We address the limitations of prior works, \fone~\cite{micro-2021-f1} in particular, and suggest a suitable architecture for bootstrappable CKKS instances. 
Section~\ref{sec:sub_3_desirable_parameters} derived the optimality of such CKKS instances assuming that an HE accelerator can hide all the computation time within the loading time of an \evk. 
\name exploits massive parallelism innate in HE ops to satisfy that optimality requirement indeed, with enough, but not an excess of, functional units (FUs).
To achieve this, first we dissect key-switching, which appears in both HMult and HRot, and has both heavy computation and memory requirements.

\subsection{Computational breakdown of HE ops}
\label{sec:sub_4_computational_breakdown}

We first \bhl{dissect} key-switching, which appears in both \HMult and \HRot, the two dominant HE ops for bootstrapping and general HE workloads.
Fig.~\ref{fig:key_switch} shows the computational flow of key-switching, and Fig.~\ref{fig:complexity} shows the corresponding computational complexity breakdown.
We focus on three functions, \emph{NTT}, \emph{iNTT}, and \emph{BConv}, which take up most of the computation.

\noindent \textbf{Number Theoretic Transform (NTT)}: A polynomial mult between polynomials in $R_Q$ translates to a negacyclic convolution of their coefficients.
NTT is a variant of the Discrete Fourier Transform (DFT) in $R_Q$.
Similar to DFT, NTT transforms the convolution between two sets of coefficients into an element-wise mult, while inverse NTT (iNTT) is applied to obtain the final result as shown below ($\otimes$ meaning element-wise mult):
\begin{equation*}
\begin{split}
   &a_1(X) \cdot a_2(X) = \text{iNTT}(\text{NTT}(a_1(X)) \otimes \text{NTT}(a_2(X)))
 \end{split}
\end{equation*}
\vspace{-0.15in}

By applying the well-known Fast Fourier Transform (FFT) algorithms~\cite{cooley-tukey}, the computational complexity of (i)NTT is reduced from $\mathcal{O}(N^2)$ to $\mathcal{O} (N\log{}N)$.
This strategy divides the computation into $\log N$ stages, where $N$ data elements are paired into $\sfrac{N}{2}$ pairs in a strided manner and butterfly operations are applied to each pair per stage. 
The stride value changes every stage.
Butterfly operations in (i)NTT are as follows:
\begin{equation*}
\begin{split}
&\text{Butterfly}_\text{NTT}(X, Y, W)\!\rightarrow\!X^{\prime}\!=\!X\!+\!W\!\cdot\!Y, Y^{\prime}\!=\!X\!-\!W\!\cdot\!Y \\
&\text{Butterfly}_\text{iNTT}(X, Y, W^{-1})\!\rightarrow\!X^{\prime}\!=\!X\!+\!Y, Y^{\prime}\!=\!(X\!-\!Y)\!\cdot\!W^{-1}
\end{split}
\end{equation*}
\vspace{-0.05in}

\noindent where $W$ (a \emph{twiddle factor}) is an odd power (up to $2N-1$) of the primitive $2N$-th root of unity $\xi$.
In total, $N$ twiddle factors are needed \emph{per prime modulus}.
NTT can be applied concurrently to each residue polynomial (in $R_{q_i}$) in a \ct.

\noindent \textbf{Base Conversion (BConv)}: 
BConv~\cite{sac-2016-behz} converts a set of residue polynomials to another set whose prime moduli are different from the former.
A \ct at level $\ell$ has two polynomials, with each consisting of $(\ell + 1)$ residue polynomials corresponding to prime moduli $\{q_0, ..., q_{\ell}\}$.
We denote this modulus set as $C_{\ell}$, called the polynomial's base or \emph{base} in short.

BConv is required in key-switching to match the base of a \ct with an \evk on base $B \cup C_{\ell}$ where $B=\{p_0, ..., p_{k-1}\}$.
BConv from $C_{\ell}$ to $B$ is performed on \cts, as expressed in Eq.~\ref{eq:BConv}, where $\hat{q_j}\!=\!\prod_{i \neq j}q_i$ for $q_i\!\in\! C_{\ell}$.
Likewise, BConv from $B$ to $C_{\ell}$ is performed after multiplying \ct by \evk.
\begin{equation}
    \underset{C_{\ell}\rightarrow B}{\text{BConv}}([a(X)]_{C_{\ell}})\! =\! \Bigg\{\! \Bigg[\sum_{j=0}^{\ell}\underbrace{[[a(X)]_{q_j}\! \cdot\! \hat{q_j}^{-1}]_{q_j}}_{\text{(1)}}\! \cdot \hat{q_j} \Bigg]_{p_i}\! \Bigg\}_{0 \leq i < k}
    \hspace{-0.2in}
    \label{eq:BConv}
\end{equation}
\vspace{-0.07in}

Because BConv cannot be performed on polynomials after NTT (i.e., they are in the \emph{NTT domain}), iNTT is performed to bring the polynomials back to the \emph{RNS domain}.
\name keeps polynomials in the NTT domain by default and brings them back to the RNS domain only for BConv.
Thus, a sequence of $\text{iNTT}\! \rightarrow\! \text{BConv}\! \rightarrow\! \text{NTT}$ is a common pattern in CKKS.

\subsection{Limitations in prior works and the balanced design of \name}
\label{sec:limitations_in_prior_works}

Prior HE acceleration studies~\cite{micro-2021-f1, asplos-2020-heax, hpca-2019-roy, hpca-2021-cheetah} identified (i)NTT as the paramount acceleration target and placed multiple \emph{NTT units} (NTTUs) that can perform both Butterfly\textsubscript{NTT} and Butterfly\textsubscript{iNTT}.
\fone~\cite{micro-2021-f1} in particular populated numerous NTTUs with ``the more the better'' approach, provisioning 14,336 NTTUs even for a small HE parameter set with $N \! = \!2^{14}$.
Such an approach was viable because, under 
the small parameter sets, all \ct, \evk, and temporal data could reside on-chip, especially with proper compiler support.

However, we observe that such massive use of NTTUs is wasteful in bootstrappable CKKS instances, where the off-chip memory bandwidth becomes the main determinant of the overall performance. 
The FHE-optimized parameters cause a quadratic increase in \ct, \evk, and the temporal data (e.g., 64$\times$ when moving from $2^{14}$ to $2^{17}$ of $N$). This makes it impossible for these components to be located on-chip, especially considering that most prior custom hardware works only take into account the max \dnum case.

We instead analyze how many fully-pipelined NTTUs an HE accelerator requires to finish \HMult or \HRot within the \evk loading time with our target CKKS instances.
We define the minimum required number of NTTUs (min$_\text{NTTU}$) as $\frac{\text{\# of butterflies per HE op}}{\text{operating frequency}}/$ $\frac{\text{size of an }\mathbf{evk}}{\text{main-memory bandwidth}}$.
When we assume a nominal operating frequency of 1.2GHz for NTTUs considering prior works~\cite{micro-2021-a100, hcs-2021-graphcore, isca-2021-tpuv4i} in 7nm process nodes, and HBM with an aggregate bandwidth of 1TB/s, min$_\text{NTTU}$ is defined as shown below:
\begin{equation}
\begin{split}
    \text{min}_{\text{NTTU}}\! &=\! \frac{(\mathtt{dnum}\! +\! 2)\! \cdot\! (k\! +\!\ell\! +\! 1)\! \cdot\! \frac{1}{2}N \log N / (1.2\text{GHz})}{2\! \cdot\! \mathtt{dnum}\! \cdot\! (k\! +\!\ell\! +\! 1)\! \cdot\! N\! \cdot\! 8\text{B} / (1\text{TB}/\text{s}) }
\end{split}
\label{eq:min_nttu}
\end{equation}
\vspace{-0.1in}

\noindent
The value of $\text{min}_{\text{NTTU}}$ is maximized when \dnum is 1. For $N=2^{17}$, the value is 1,328.
We utilize 2,048 NTTUs in \name to provide some margin for other operations.

In addition to (i)NTT, the importance of BConv grows as small \dnums are used.
The computational complexity of BConv in key-switching is proportional to ($1\!+\!\frac{2}{\mathtt{dnum}}$).
As a result, the relative computational complexity of BConv, which is 12\% at $\mathtt{dnum}\!=\!\text{max}$, increases to 34\% at $\mathtt{dnum}\!=\!1$ (see Fig.~\ref{fig:key_switch_and_complexity}(b)).
Prior works mainly targeted $\mathtt{dnum}\!=\!\text{max}$, focusing on the acceleration of (i)NTT.
We propose a novel \emph{BConv unit} (BConvU) to handle the increased significance of BConv,
whose details are described later in 
Section~\ref{sec:sub_5_baseconv}.

\subsection{\name organization exploiting data parallelism}
\label{sec:sub_4_parallelism_HE_function}

\begin{figure}[tb!] 
    \begin{center}
    \includegraphics[width=0.8\columnwidth]{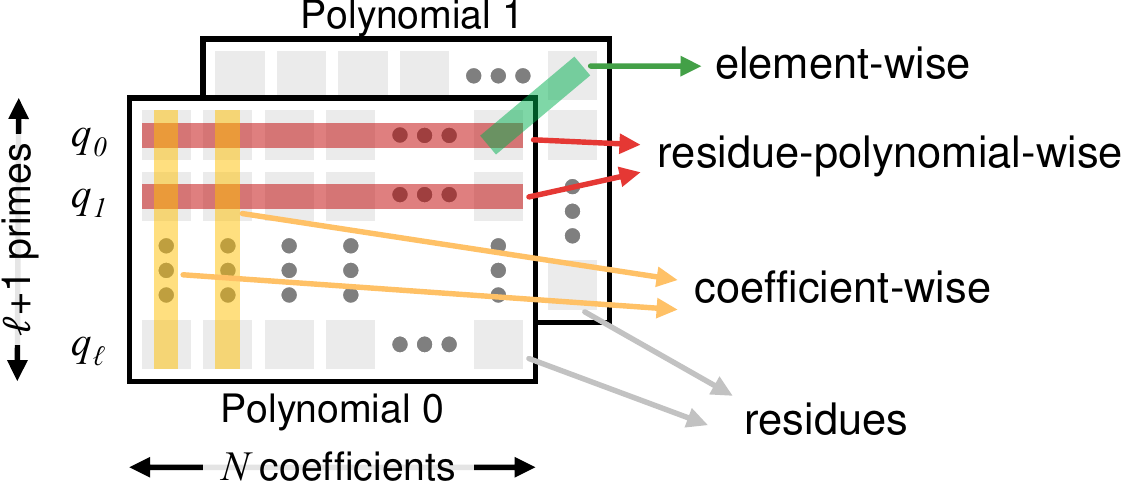}
    \end{center}
    \vspace{-0.1in}
    \caption{Data access patterns in HE functions.\label{fig:parallelism}}
    \vspace{-0.1in}
\end{figure}

We can categorize primary HE functions into three groups according to their data access patterns (see Fig.~\ref{fig:parallelism}).
Residue-polynomial-wise functions, the (i)NTT and automorphism functions, involve all $N$ residues in a residue polynomial to produce an output.
Coefficient-wise functions (e.g., BConv) involve all $(\ell\!+\!1)$ residues of a single coefficient to produce an output residue.
Element-wise functions such as CMult and PMult only involve residues on the same position over multiple polynomials.

We can exploit two types of data parallelism, \emph{residue-polynomial-level parallelism} (rPLP) and \emph{coefficient-level parallelism} (CLP), when parallelizing an HE op with multiple processing elements (PEs).
rPLP can be exploited by distributing $(\ell\!+\!1)$ residue polynomials and CLP can be by distributing $N$ coefficients to many PEs.
Prior works including \fone mostly exploited rPLP as prime-wise modularization is apparently possible.

When the data access pattern and the type of the parallelism being exploited are not aligned, data exchanges between PEs occur, resulting in global wire communication which has poorly scaled over technology generations~\cite{ieee-2001-wire}.
For the sequence of $\text{iNTT}\! \rightarrow\! \text{BConv}\! \rightarrow\! \text{NTT}$ in key-switching, CLP will incur data exchanges for (i)NTT and rPLP will incur data exchanges for BConv. The total size of the transferred data is identical at $(k\!+\!\ell\!+\!1)N$.
Thus, there is no clear winner between the two types of parallelism in terms of data exchanges.
However, exploiting rPLP is limited in terms of the degree of parallelism due to the fluctuating multiplicative level $\ell$ as an FHE application is executed. This also complicates a fair distribution of jobs among PEs.

Instead, we use CLP in \name.
As $N$ is fixed throughout the running of an HE application, we decide on a fixed data distribution methodology, where the residues of a polynomial with the same coefficient index are allocated to the same PE.
Then, coefficient-wise and element-wise functions are parallelized without inter-PE data exchanges; only (i)NTT and the automorphism incur inter-PE data exchanges, with the communication pattern predetermined by the fixed data distribution.

We place 2,048 PEs (Eq.~\ref{eq:min_nttu}) in \name.
Each PE has an NTTU, a BConvU, a modular adder (ModAdd) and a multiplier (ModMult) for element-wise functions, as well as an SRAM scratchpad.
$N\!=\!2^{17}$ residues of a residue polynomial are evenly distributed to the PEs, such that one PE handles $2^6$ residues.
Then six out of 17 (i)NTT stages can be solely computed inside a PE.
We adopt \emph{3D-NTT} to minimize the data exchanges between the PEs.
A residue polynomial is regarded as a 3D data structure of size $2^6 \times 2^5 \times 2^6$.
Then, each PE performs a sequence of $2^6$-, $2^5$-, and $2^6$-point (i)NTTs, interleaved with just two rounds of inter-PE data exchange.
Splitting (i)NTT in a more fine-grained manner requires more data exchange rounds and is thus less energy-efficient.
The automorphism function exhibits a different communication pattern from (i)NTT, involving complex data remapping (Eq.~\ref{eq:automorphism}).
Nevertheless, the data distribution methodology and NoC structure of \name efficiently handle data exchanges for both (i)NTT and the automorphism (see Section~\ref{sec:5_microarchitecture}).

%% file: 5_contribution-3.tex
\begin{figure*}[tb!] 
    \begin{center}
    \includegraphics[width=0.99\textwidth]{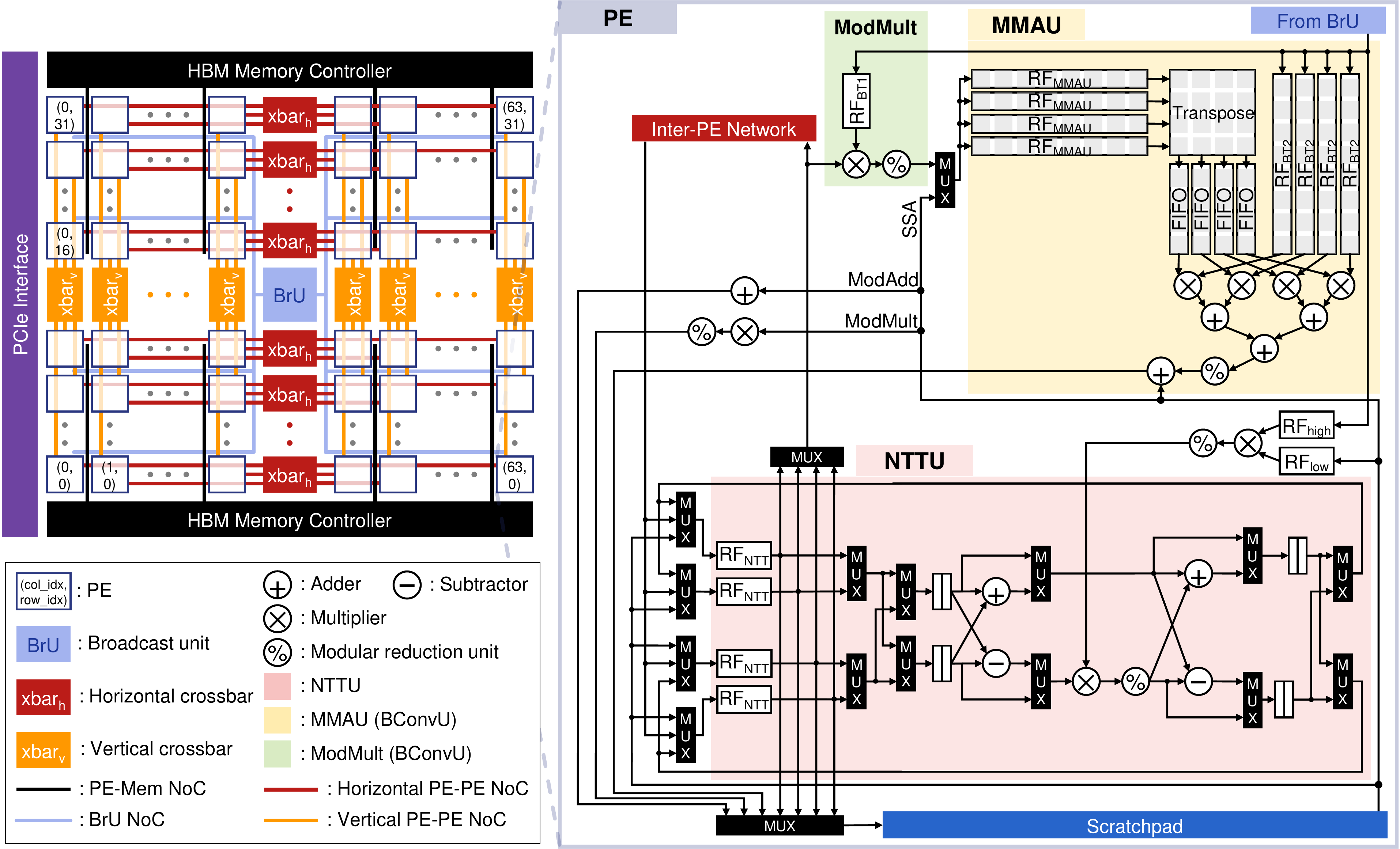}
    \end{center}
    \vspace{-0.1in}
    \caption{The overview of \name: Each PE in a grid is denoted as (column index, row index).
PEs interconnect through the PE-PE NoC composed of xbar\textsubscript{v} and xbar\textsubscript{h}.
BrU is the broadcast unit.
BrU and the main memory communicate with PEs through separate NoCs.
A PE consists of a scratchpad, an NTTU to undertake NTT/iNTT, a BConvU for BConv, a modular multiplier (ModMult), and a modular adder (ModAdd). BConvU consists of a ModMult and MMAU.}
    \vspace{-0.05in}
    \label{fig:bts_architecture}
\end{figure*}

\section{\name Microarchitecture}
\label{sec:5_microarchitecture}

We devise a massively parallel architecture that distributes PEs in a grid.
A PE consists of functional units (FUs) and an SRAM scratchpad.
An NTTU in each PE handles a portion of the residues in a residue polynomial during (i)NTT.
By exploiting CLP,
the coefficient-wise or element-wise functions
can be computed in a PE without any inter-PE data exchange.

Fig.~\ref{fig:bts_architecture} presents a high-level overview of \name.
We arrange 2,048 ($\npe$) PEs in a grid with a vertical height of 32 ($\npever$) and a horizontal width of 64 ($\npehor$).
The PEs are interconnected via dimension-wise crossbars in the form of 32$\times$32 vertical crossbars (xbar\textsubscript{v}) and 64$\times$64 horizontal crossbars (xbar\textsubscript{h}). We populate a central, constant memory, storing precomputed values including twiddle factors for (i)NTT and $\hat{q}_j, \hat{q}_j^{-1}$ for BConv.
A broadcast unit (BrU) delivers the precomputed values to the PEs at the required moments.
Memory controllers are located at the top and bottom sides, each connected to an HBM stack.
\name receives instructions and necessary data from the host via the PCIe interface.
The word size in \name is 64 bits.
Modular reduction units use Barrett reduction~\cite{eurocrypt-1986-barrett} to bring the 128-bit multiplied results back to the word size.

\subsection{Datapath for (i)NTT}
\label{sec:sub_5_datapath_NTT}

\name maps the coefficients of a polynomial to the PEs suited to 3D-NTT.
We view the $N$ residues in a residue polynomial as a $(N_x, N_y, N_z)\! =\! (\npehor, \npever, \sfrac{N}{\npe})$ cube.
Then in the RNS domain, a residue at the coefficient index $i$ (the coefficient of $X^i$) is at position $(x, y, z)$ in this cube, where $i = x + N_x \cdot y + N_x \cdot N_y \cdot z$.
We allocate residues at position $(x',y',z')_{z' \in [0,N_z)}$ of such a cube to the PE of $(x',y')$ coordinate in the PE grid.
3D-NTT is broken down into five steps in \name.
First, we conduct i) \nttz inside a single PE, which corresponds to the NTT along the z-axis of the cube.
Next, ii) data exchanges between vertically aligned PEs are executed, corresponding to $\npehor$ of yz-plane parallel transposition of residues in the cube.
iii) \ntty along the z-axis follows.
iv) Data exchanges between horizontally aligned PEs are executed, corresponding to $\npever$ of xz-plane parallel transposition of residues in the cube.
Finally, v) \nttx along the z-axis is carried out.
iNTT is performed by the reverse process of NTT.

An NTTU supports both NTT and iNTT by using logic circuits similar to~\cite{iscas-2021-multiplier, tcas-2020-vpqc, tches-2021-compact, isca-2021-pipezk}.
We employ separate register files (RF\textsubscript{NTT}s) to reuse data between (i)NTT stages.
An NTTU decomposes \nttx, \ntty, and \nttz into radix-2 NTTs.
It is fully pipelined and performs one butterfly op per clock.
An input pair is fed in, and an output pair is stored from the NTTU each cycle, provided by two pairs of RF\textsubscript{NTT}s.

We hide the time for vertical and horizontal data exchanges of 3D-NTT (steps ii) and iv)) through coarse-grained, \emph{epoch}-based pipelining.
As steps i), iii), and v) are executed with the same NTTU, we determine the length of an epoch according to the time required to perform these three steps 
($ \frac{N\log\! N}{2\cdot \npe}$ cycles).
Within the $r$-th epoch, we time-multiplex i) of $(r\!+\!2)$-th, iii) of $r$-th, and v) of the $(r\!-\!2)$-th residue polynomials, while exchanging ii) of $(r\!+\!1)$-th and iv) of the $(r\!-\!1)$-th residue polynomials concurrently.
Concurrent data exchanges are enabled by separate vertical (ii)) and horizontal (iv)) NoCs.
Thus, (i)NTT of a single residue polynomial finishes every epoch.

A single (i)NTT on a residue polynomial requires $N$ different twiddle factors. Because each prime modulus needs different twiddle factors, the sizes of the twiddle factors for (i)NTT on a ciphertext reach dozens of MBs for our target CKKS instances.
We reduce the storage for the twiddle factors by decomposing them by means of on-the-fly twiddling (OT)~\cite{iiswc-2020-ntt}.
OT replaces the $N$-sized precomputed twiddle-factor table with two tables: a \emph{higher-digit table} of $\xi_{2N}^{mj}$ where $1$$\leq\!j$$< \sfrac{(\!N\!-\!1\!)}{m}$, and a \emph{lower-digit table} of $\xi_{2N}^i$ where $1$$\leq$$i$$<$$m$. 
We can compose any twiddle factor $\xi_{2N}^k$ by multiplying two twiddle factors $\xi_{2N}^i$ and $\xi_{2N}^{mj}$ that satisfy $k\!=\!mj\!+\!i$.
OT reduces the memory usage by $2/m$.
\name stores the lower-digit tables of prime moduli in PEs (each PE having different entries) while storing the higher-digit tables in the \BrU (all PEs sharing the entries).
The BrU broadcasts a higher-digit table for a prime modulus to PEs for every (i)NTT epoch.

\subsection{Base Conversion Unit (BConvU)}
\label{sec:sub_5_baseconv}

BConv consists of two parts.
The first part multiplies residue polynomials with $\left[\hat{q_j}^{-1}\right]_{q_j}$ and the second part does this with $\left[\hat{q_j}\right]_{p_i}$ and accumulates them.
It is the second part that exhibits the coefficient-wise access pattern because it accumulates residues at the same coefficient index in all residue polynomials.

A BConv unit (BConvU) with a modular multiplier (ModMult) for the first part and a modular multiply-accumulate unit (MMAU) for the second part is placed in each PE.
BConv strongly depends on the preceding iNTT (see Fig.~\ref{fig:key_switch_and_complexity}).
Because iNTT is a residue-polynomial-wise function, whereas the second part of BConv is a coefficient-wise function, the MMAU must wait until iNTT is finished on all residue polynomials. 
We mitigate this by partially overlapping iNTT and BConv. 
We modify the right-hand side of Eq.~\ref{eq:BConv} as follows:
\begin{equation}
\bigg\{\sum_{j_1=0}^{(\ell+1)/\lsub-1}\!\bigg[\!\sum_{j_2=j_1\times \lsub}^{(j_1+1)\times \lsub-1}\![[a(X)]_{j_2}\!\cdot \!\hat{q}_{j_2}^{-1}]_{q_{j_2}}\!\cdot\! \hat{q}_{j_2}\!\bigg]_{p_i}\!\bigg\}_{0\leq i < k} \hspace{-0.2in}
\label{eq:BConv_modified}
\end{equation}
\vspace{-0.05in}

This modification enables the second part to start when the preceding iNTT and the first part of BConv are finished on $l$\textsubscript{sub}($=\!4$ in \name) residue polynomials and stored in RF\textsubscript{MMAU}.
The MMAU computes the corresponding partial sum (the inner sum of Eq.~\ref{eq:BConv_modified}), and accumulates this result with the previous results (the outer sum), which are loaded from and stored on to a scratchpad inducing a read and write every cycle.
Temporal registers and FIFO minimize the bandwidth pressure on RF\textsubscript{MMAU} and transpose the data for the correct orientation to feed $l$\textsubscript{sub} lanes into the MMAU.
The precomputed values of $[\hat{q_j}^{-1}]_{q_j}$ and $[\hat{q_j}]_{p_i}$ (\emph{BConv tables})
are respectively loaded into the dedicated $\text{RF}_\text{BT1}$ and $\text{RF}_\text{BT2}$ from the \BrU when needed.

We also leverage the MMAU for other operations.
Subtraction, $\sfrac{1}{P}$ scaling, and $d0$/$d1$ addition at the end of key-switching (Fig.~\ref{fig:key_switch_and_complexity}) can be expressed as $[d2^{\prime}.ax]_{Q_{\ell}}\! \times\! (1/P) + [d2^{\prime}.ax]_{P\rightarrow Q_{\ell}}\! \times\! (-1/P) + d1\! \times\! 1 + 0\! \times\! 0$; thus, we fuse these three operations to be computed on the MMAU.
We refer to this fusion as subtraction-scaling-addition (SSA).

\subsection{Scratchpad}
\label{sec:sub_5_scrartchpad_design}

The per-PE scratchpad has three purposes. First, it stores the temporary data generated during the course of the HE ops.
The size of the temporal data during key-switching can be large (e.g., a single (i)NTT or BConv can produce 28MB at $\ell\!+1\!=\!28$, $N\!=\!2^{17}$). 
If such data does not reside on-chip, the additional off-chip access would cause severe performance degradation.

Second, the \scratchpad also stores the prefetched \evk. 
To hide the latency of the \evk load time, it must be prefetched beforehand.
As \evk is not consumed immediately after being loaded on-chip, it takes up a portion of the \scratchpad.

Third, the \scratchpad functions as a cache for \cts, controlled explicitly by software (SW caching).
\cts often show high temporal locality during a sequence of HE ops.
For instance, during bootstrapping, a \ct is commonly subjected to multiple HRots.
Moreover, as HE ops form a deterministic computational flow and the granularity of cache management is as large as a \ct, SW control is manageable.

The scratchpad bandwidth demand of the BConvU is high (as later detailed in Fig.~\ref{fig:gantt-chart}) due to the accesses involved when updating the partial sums.
Considering that the partial sum size is only proportional to $k$ in Eq.~\ref{eq:BConv_modified} and is loaded $(\ell\!+\!1) / \lsub$ times, the bandwidth pressure can be relieved by increasing $\lsub$.
However, this would also require an increase in the number of lanes in the MMAU (and hence the size of $\text{RF}_\text{MMAU}$), resulting in a trade-off.

\subsection{Network-on-Chip (NoC) design}
\label{sec:sub_5_NoC_design}

\name has three types of on-chip communication: 
1) off-chip memory traffic to the PEs (PE-Mem NoC),
2) the distribution of precomputed constants to PEs (BrU NoC), and
3) inter-PE data exchanges for (i)NTT and the automorphism (PE-PE NoC).
\name has a large number of nodes  (over 2k endpoints) and requires a high bandwidth.
Given the unique communication characteristics of each type of on-chip communication, \name provides three separate NoCs instead of sharing a single NoC to enable deterministic communication while minimizing the NoC overhead. 

\noindent \textbf{PE-Mem NoC}: Because data is distributed evenly across the PEs, the off-chip memory (i.e., HBM2e~\cite{jedec-2021-hbm2}) is placed on the top and bottom and each HBM only needs to communicate with half of the PEs placed nearby.
The PE grid placement is exploited by separating the PEs into 32 regions and connecting each HBM pseudo-channel only to a single PE region.
An HBM2e stack supports 16 pseudo-channels~\cite{micron-hbm2e} and thus the upper half of the PEs \bhl{has} 16 regions while the lower half also has 16 regions, with each region consisting of 64 \bhl{PEs.}

\noindent \textbf{BrU NoC}: BrU data is globally shared by all PEs and broadcast to all PEs.
Given the large number of PEs, the BrU is organized hierarchically with 128 \emph{local BrUs}. Each \emph{local BrU} provides higher-digit tables of twiddle factors and BConv tables to 16 PEs.
The global BrU is loaded with all precomputed values before an HE application starts and sends data to the local BrUs that serve as temporary storage/repeaters.

\noindent \textbf{PE-PE NoC}:
The PE-PE NoC requires support for the highest bandwidth due to  the data exchanges necessary between the PEs.
The communication pattern is \emph{symmetric} (i.e., each PE sends and receives the same amount of data), and a single PE is not oversubscribed.
In addition, because the traffic pattern is known (e.g., all-to-all or a fixed, permutation traffic), the NoC can be greatly simplified.
\name implements a logical 2D flattened butterfly~\cite{micro-2007-fb,sc-2009-hyperx} given that communication to other PEs within each row and within each column is limited.
However, instead of having a router at each PE, a single ``router'' xbar\textsubscript{h} (respectively, xbar\textsubscript{v}) is shared by all PEs within each row (column); it is placed in the center of each row (column) and used for horizontal (vertical) data exchange steps of (i)NTT (steps ii), iv)).
Each xbar\textsubscript{h} (xbar\textsubscript{v}) does not require any allocation because the traffic pattern is known ahead of time and can be scheduled through pre-determined arbitration.

\subsection{Automorphism}
\label{sec:sub_5_automorphism}

We identify that \name can handle the automorphism for HRots efficiently.
All residues mapped to a single PE always move to another single destination PE under the \name' PE-coefficient mapping scheme;
i.e., the inter-PE communication of the automorphism exhibits a permutation pattern.
A PE of the $(x^{\prime}, y^{\prime})$ PE-grid coordinate holds the residues at positions $(x^{\prime}, y^{\prime}, z^{\prime})_{z^{\prime} \in [0,N_z)}$, corresponding to coefficient indices $i\! =\! x^{\prime}\! +\! N_x \! \cdot \! y^{\prime}\! +\! N_x \! \cdot \! N_y \! \cdot \! z^{\prime}$ (Section~\ref{sec:sub_5_datapath_NTT}).
$i$s in binary format only differ in the higher
bit-field ($N_x \! \cdot \! N_y \! \cdot \! z^{\prime}$), meaning that the automorphism destination indices ($i\cdot 5^r$'s in Eq.~\ref{eq:automorphism}) also only differ in the higher bit-field; the residues are mapped to the same destination PE corresponding to the lower bit-field ($x^{\prime\prime}\! +\! N_x \cdot y^{\prime\prime}$).

We can decompose such a permutation pattern into three steps to fit the PE-PE NoC structure of \name:
intra-PE permutation (z-axis), vertical permutation (y-axis), and horizontal permutation (x-axis).
Each step gradually updates the $i$s to $i \cdot 5^r$s from higher to lower bit-fields.
The intra-PE permutation process does not use the NoC.
The vertical/horizontal permutations can be handled by xbar\textsubscript{v}/xbar\textsubscript{h}.
The PE-PE NoC can support an arbitrary HRot with any rotation amount ($r$) without data contention, whose property is similar to that of 3D-NTT.

%% file: 6_evaluation.tex
\section{Evaluation}
\label{sec:6_evalutation}

\subsection{Hardware modeling of \name}
\label{sec:sub_6_hardware_modeling}

\begin{table}[tb!]
  \centering
  \caption{The area and the peak power of components in \name.\label{tab:area_power}}
  \begin{tabular}{l@{\hspace{0.05in}}|@{\hspace{0.1in}}r@{\hspace{0.01in}}l@{\hspace{0.02in}}r@{\hspace{0.05in}}l@{\hspace{0.02in}}r}
  \toprule
  & Area && Power && Freq \\
  Component          &  ($\mu$m\textsuperscript{2}) &&  (mW)                               &&  (GHz) \\
  \midrule
  Scratchpad SRAM    & 114,724                          && 9.86                                     && 1.2        \\
  RFs                & 12,479                           && 2.29                                     && Various    \\
  NTTU               & 9,501                            && 12.17                                    && 1.2        \\
  ModMult (BConvU)   & 4,070                            && 0.56                                     && 0.3        \\
  MMAU (BConvU)      & 9,511                            && 8.42                                     && 1.2        \\
  Exchange unit      & 421                              && 1.03                                     && 1.2        \\
  ModMult            & 3,833                            && 1.35                                     && 0.6        \\
  ModAdd             & 325                              && 0.08                                     && 0.6        \\
  \midrule
  \textbf{1 PE}      & \textbf{154,863}                 && \textbf{35.75}                           && -          \\
  \bottomrule
  \toprule
  & Area && Power && Freq \\
  Component          &  (mm\textsuperscript{2})     &&  (W)                                && (GHz) \\
  \midrule
  2048 PEs           & 317.2\enspace                    && 73.21                                    && -          \\
  Inter-PE NoC       & 3.06                             && 45.93                                    && 1.2        \\
  Global BrU + NoC   & 0.42                             && 0.10                                     && 0.6        \\
  128 local BrUs     & 3.69                             && 0.04                                     && 0.6        \\
  HBM2e NoC            & 0.10                             && 6.81                                     && 1.2        \\
  2 HBM2e stacks & 29.6\enspace & \cite{isca-2021-tpuv4i} & 31.76 & \cite{micro-2017-finegrainedDRAM} & -      \\
  PCIe5x16 interface & 19.6\enspace & \cite{isca-2021-tpuv4i} & 5.37 & \cite{cicc-2020-pcie5}            & -      \\
  \midrule
  \textbf{Total}     & \textbf{373.6}\enspace           && \textbf{163.2}\enspace                   && -          \\
  \bottomrule
  \end{tabular}
  \vspace{-0.1in}
\end{table}

We used the ASAP7~\cite{mj-2016-asap7, mse-2017-asap7collateral} design library to synthesize the logic units and datapath components in a 7nm technology node.
We simulated the RFs and \scratchpads using FinCACTI~\cite{isvlsi-2014-fincacti} due to the absence of a public 7nm memory compiler.
We updated the analytic models and technology constants of FinCACTI to match ASAP7 and the IRDS roadmap~\cite{whitepaper-2018-irds}.
We validated the RTL synthesis and SRAM simulation results against published information~\cite{isscc-2017-7nm-sram, isscc-2018-7nm-sram-euv, iedm-2017-intel10nm, iedm-2016-tsmc7nm, iedm-2017-gf7nm, isca-2021-tpuv4i, vlsit-2018-samsung}.

\name uses single-ported 128-bit wide 1.2GHz SRAMs for the \scratchpads, providing a total capacity of 512MB and a bandwidth of 38.4TB/s chip-wide.
RFs are implemented in single-ported SRAMs with variable sizes, port widths, and operating frequencies following the requirements of the FUs. 
22MBs of RFs are used chip-wide, providing 292TB/s.
Crossbars in the PE-PE NoC have 12-bit wide ports and run at 1.2GHz, providing a bisection bandwidth of 3.6TB/s.
The NoC wires are routed over other components~\cite{tcadics-2012-crossbar100}.
We analyzed the cost of wires and crossbars using FinCACTI and prior works~\cite{ted-2002-repeater, whitepaper-2018-irds, intel-2008-wire45nm, tcadics-2012-crossbar100}.
Two HBM2e stacks are used~\cite{jedec-2021-hbm2}, but with a modest 11\% speedup assumed, considering the latest technology~\cite{jedec-2022-hbm3}.
The peak power and area estimation results are shown in Table~\ref{tab:area_power}.
\name is 373.6mm\textsuperscript{2} in size and consumes up to 163.2W of power.

\subsection{Experimental setup}
\label{sec:sub_6_experimental_setup}

We developed a cycle-level simulator to model the compute capability, latency, and bandwidth of the FUs and the memory components composing \name. When an HE op is called, the simulator converts the op into a computational graph with primary HE functions. Based on the derived computation and data dependencies, the simulator schedules functions and data loads in epoch granularity while minimizing the temporary data hold time. Utilization rates are also collected and combined with the power model to calculate the energy.
The scratchpad space is prioritized in the order of the temporary data, prefetched \evk, and finally, \ct caching with an LRU policy.

We measured \amormultslot as a microbenchmark and evaluated the most complex applications currently available on CKKS: logistic regression (HELR~\cite{aaai-2019-helr}), CNN inference (ResNet-20~\cite{arxiv-2021-resnet20}), and sorting~\cite{tifs-2021-sorting}.
HELR trains a binary classification model with MNIST~\cite{2012-mnist} for 30 iterations, each with a batch containing 1,024 14$\times$14-pixel images.
ResNet-20 performs homomorphic convolution, linear transform, and ReLU.
It achieves 92.43\% accuracy on CIFAR-10 classification~\cite{techreport-2009-cifar10}.
We used the channel packing method proposed in~\cite{usenixsec-2018-gazelle} to pack all of the feature map channels into a single \ct to improve the performance further.
Sorting uses a 2-way sorting network to sort $2^{14}$ data.
Because non-linear functions such as ReLU and comparisons are approximated by high-degree polynomial functions in CKKS, they consume many levels and induce hundreds of bootstrapping for ResNet-20 and sorting, respectively.

\setlength{\tabcolsep}{4.5pt}
\begin{table}[tb!]
  \centering
  \caption{The CKKS instances used for evaluation.\label{tab:parameter}}
  \begin{tabular}{c|cccccc}
  \toprule
  CKKS instance      & $N$      & $L$ & $\mathtt{dnum}$ & $\log PQ$ & $\lambda$ & Temp data \\
  \midrule
  \nameone & $2^{17}$ & 27  & 1               & 3090      & 133.4     & 183MB     \\
  \nametwo & $2^{17}$ & 39  & 2               & 3210      & 128.7     & 304MB     \\
  \namethree & $2^{17}$ & 44  & 3               & 3160      & 130.8     & 365MB     \\
  \bottomrule
  \end{tabular}
  \vspace{-0.1in}
\end{table}
\setlength{\tabcolsep}{6pt}

We compared \name with the state-of-the-art implementations on a CPU (\lattigo~\cite{github-lattigo}), a GPU (\hundredx~\cite{tches-2021-100x}), and an ASIC (\fone~\cite{micro-2021-f1}) for \amormultslot and HELR.
We ran \lattigo on a system with an Intel Skylake CPU (Xeon Platinum 8160) and 256GB of DDR4-2666 memory.
We used the 128b-secure CKKS instance preset of \lattigo and newly implemented HELR on \lattigo.
For \hundredx and \fone, the execution times reported in each paper were used.
\hundredx~\cite{tches-2021-100x} used NVIDIA V100~\cite{techreport-2017-v100} for the evaluation.
We also compared \name with \foneplus, whose execution times are optimistically scaled from \fone to have the same area as \name at 7nm~\cite{iedm-2017-gf7nm}.
For other applications, we compared \name with reported multi-threaded CPU performance from each paper due to the absence of available implementations.
We used the CKKS instances shown in Table~\ref{tab:parameter} to evaluate \name.
They all have the same degree and satisfy 128b security but use different values of $L$ and \dnum.
As \dnum and $L$ increase, the temporary data increases, requiring more scratchpad space.

\subsection{Performance and efficiency of \name}
\label{sec:sub_6_application_performance}

\begin{figure}[tb!] 
    \centering
    \includegraphics[width=0.88\columnwidth]{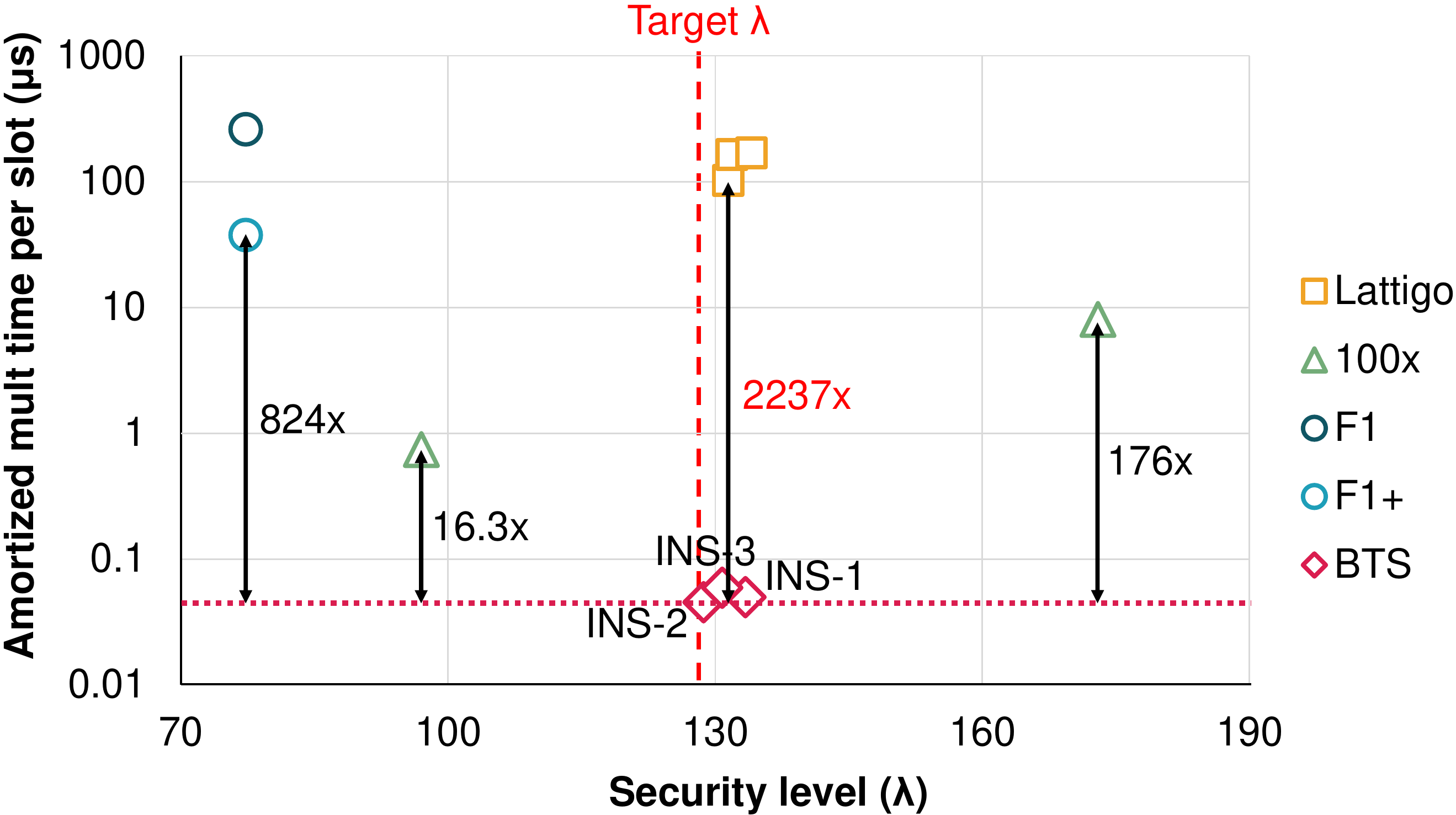}
    \vspace{-0.1in}
    \caption{Comparison of the \amormultslot between \name and other prior works of \texttt{Lattigo}~\cite{github-lattigo}, \texttt{100x}~\cite{tches-2021-100x}, and \texttt{F1}~\cite{micro-2021-f1}. \texttt{F1+} is a scaled-up version of \texttt{F1}. \texttt{INS-x} denotes the CKKS instances used for \name, specified in Table~\ref{tab:parameter}.\label{fig:amortized_mult_comparison}}
    \vspace{-0.05in}
\end{figure}

\noindent
\textbf{Amortized mult time per slot:}
\name outperforms the state-of-the-art CPU/GPU/ASIC implementations by tens to thousands of times in terms of the throughput of \HMult.
Fig.~\ref{fig:amortized_mult_comparison} shows the \amormultslot values of \lattigo, \hundredx, \fone, \foneplus and \name.
The best \amormultslot is achieved with \nametwo at 45.5ns, 2,237$\times$ better than \lattigo.
\fone is even 2.5$\times$ slower than \lattigo; this occurs because \fone only supports single-slot bootstrapping.\footnote{We call a \ct sparsely-packed if its corresponding message occupies far fewer slots compared to the maximum number of available ($\sfrac{N}{2}$).
Bootstrapping a sparsely-packed \ct reduces the computational complexity and consumes fewer levels~\cite{eurocrypt-2019-improved}.
In an extreme case using a single-slot, such an effect is maximized.
F1 only supports single-slot bootstrapping due to the lack of multiplicative levels, as it targets support of small parameter sets.}
\foneplus is better but shows 824$\times$ lower performance than \name.
\amormultslot of \hundredx is 743ns, reporting the best performance among prior works.
However, this is for a 97b-secure parameter set; when using a 173b-secure CKKS instance, \hundredx reported a 8$\mu$s \amormultslot.

\begin{figure}[tb!] 
    \begin{center}
    \includegraphics[width=0.98\columnwidth]{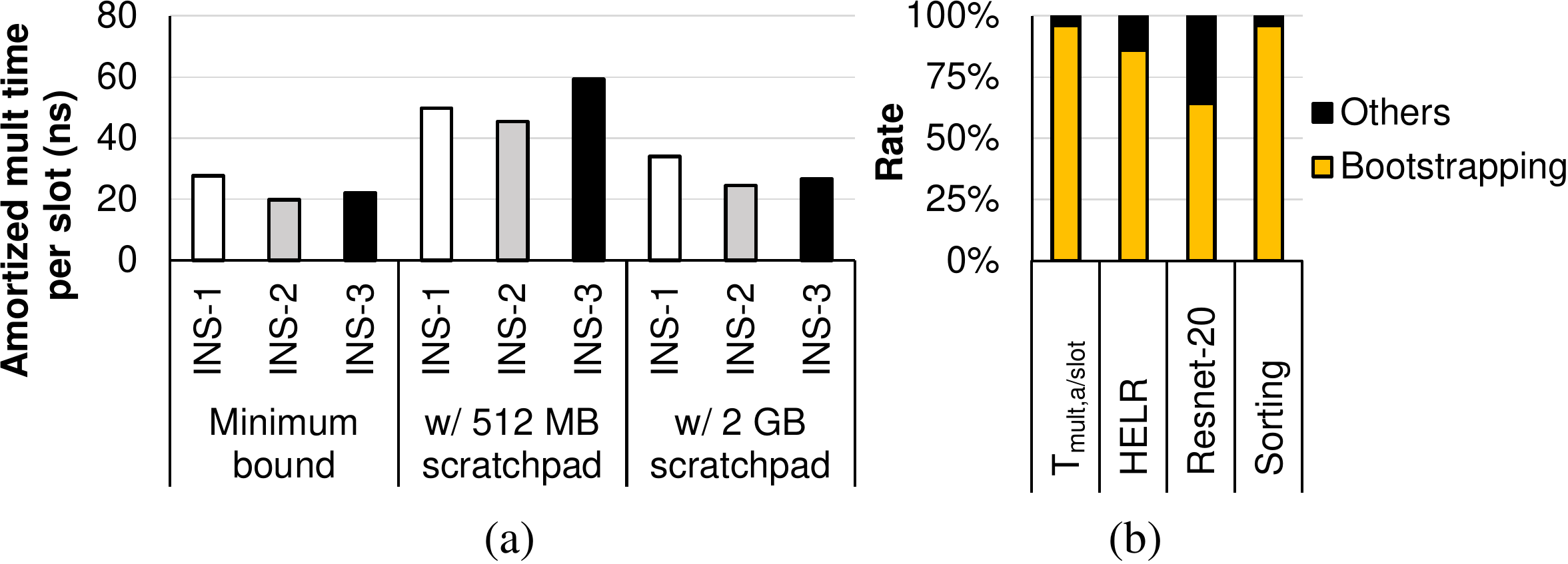}
    \end{center}
    \vspace{-0.1in}
    \caption{(a) Comparison of the minimum bound of \amormultslot (Section~\ref{sec:3_contribution1}) and the actual \amormultslot using scratchpads of 512MB and 2GB for \texttt{INS-x}, and (b) the portion of the bootstrapping time for each application on \texttt{INS-1}.}
    \vspace{-0.05in}
    \label{fig:theoretical}
\end{figure}

The performance of \namex is higher than the minimum bound performance shown in Fig.~\ref{fig:target_parameter} because \cts are not always on the scratchpad with limited capacity.
Fig.~\ref{fig:theoretical}(a) shows the minimum and actual \amormultslot using 512MB and 2GB of scratchpad for \namex.
\nametwo always performs the best.
\nameone performs better than \namethree with a 512MB scratchpad because the former requires less temporary data, leading to a higher hit rate for \cts.
With an enough (albeit not practical) scratchpad capacity of 2GB, \cts mostly hit, reaching a performance close to the minimum.

\noindent
\textbf{Logistic regression:}
Table~\ref{tab:logistic_regression} reports the average training time per iteration in HELR.
Due to the limited parameter set \fone supports,
\fone only reported the HELR training time for a single iteration with 256 images, which does not require bootstrapping but is not enough for training.
We estimated \fone's end-to-end HELR performance by assuming that 1024 images in a batch are trained over four iterations, with $14 \! \times \! 14 = \! 196$ single-slot bootstrapping applied, ignoring
the cost of packing/unpacking \cts for bootstrapping (giving favor to F1).
The execution time with \nametwo achieves 28.4ms, 1,306$\times$, 27$\times$ and 5.2$\times$ better than \lattigo, \hundredx and \foneplus, respectively.

\setlength{\tabcolsep}{2pt}
\begin{table}[tb!]
  \centering
  \caption{Comparison of performance between BTS and other prior works~\cite{github-lattigo,tches-2021-100x,micro-2021-f1} for logistic regression training~\cite{aaai-2019-helr}.\label{tab:logistic_regression}}
  \begin{tabular}{lccccccc}
  \toprule
      & \texttt{Lattigo} & \texttt{100x} & \texttt{F1} & \texttt{F1+} & \texttt{INS-1} & \texttt{INS-2} & \texttt{INS-3} \\
  \midrule
      Time (ms) & 37,050 & 775 & 1,024 & 148 & 39.9 & 28.4 & 43.5 \\
      Speedup & 1$\times$ & 48$\times$ & 36$\times$ & 250$\times$ & 929$\times$ & 1,306$\times$ & 852$\times$ \\
  \bottomrule
  \end{tabular}
  \vspace{-0.05in}
\end{table}
\setlength{\tabcolsep}{6pt}

\setlength{\tabcolsep}{3pt}
\begin{table}[tb!]
  \centering
  \caption{Evaluating \name for ResNet-20 inference~\cite{arxiv-2021-resnet20} and sorting~\cite{tifs-2021-sorting}.}
  \begin{tabular}{lcccc}
  \toprule
                                           & CPU & \texttt{INS-1} & \texttt{INS-2} & \texttt{INS-3} \\
  \midrule
  ResNet-20 execution time (s)                       & 10,602                           & 1.91           & 2.02           & 3.09            \\
  Speedup (vs. \cite{arxiv-2021-resnet20}) & 1$\times$    & 5,556$\times$  & 5,240$\times$  & 3,427$\times$  \\
  \# of bootstrapping & - & 53 & 22 & 19 \\
  \midrule
    Sorting execution time (s)                       & 23,066                             & 15.6           & 18.8           & 25.2            \\
  Speedup (vs. \cite{tifs-2021-sorting}) & 1$\times$      & 1,482$\times$  & 1,226$\times$  & 915$\times$  \\
    \# of bootstrapping & - & 521 & 306 & 229 \\
  \bottomrule
  \end{tabular}
  \label{tab:resnet-20}
  \vspace{-0.05in}
\end{table}
\setlength{\tabcolsep}{6pt}

\noindent
\textbf{ResNet-20 and sorting:}
\name performs up to 5,556$\times$ and 1,482$\times$ faster over the prior works, \cite{arxiv-2021-resnet20} and \cite{tifs-2021-sorting} (see Table~\ref{tab:resnet-20}).
For ResNet-20, \nameone without channel packing shows a 311$\times$ speedup.
By adopting the channel-packing method~\cite{usenixsec-2018-gazelle} exploiting the abundant slots of our target CKKS instances,
we reduced the working set and improved the throughput, resulting in an additional 17.8$\times$ performance gain and achieving 1.91s of ResNet-20 inference latency on an encrypted image.

Although \name provides a speedup of more than three orders of magnitude for the most complex applications, these applications still do not fully utilize all $2^{16}$ slots due to the small problem size.
We anticipate the relative speedup of \name to improve even further when real-world applications are implemented with FHE.
For instance, an ImageNet~\cite{cvpr-2009-imagenet} image has over $2^{17}$ data, which requires multiple fully-packed \cts to encrypt.

\begin{figure*}[tb!] 
    \begin{center}
    \includegraphics[width=0.99\textwidth]{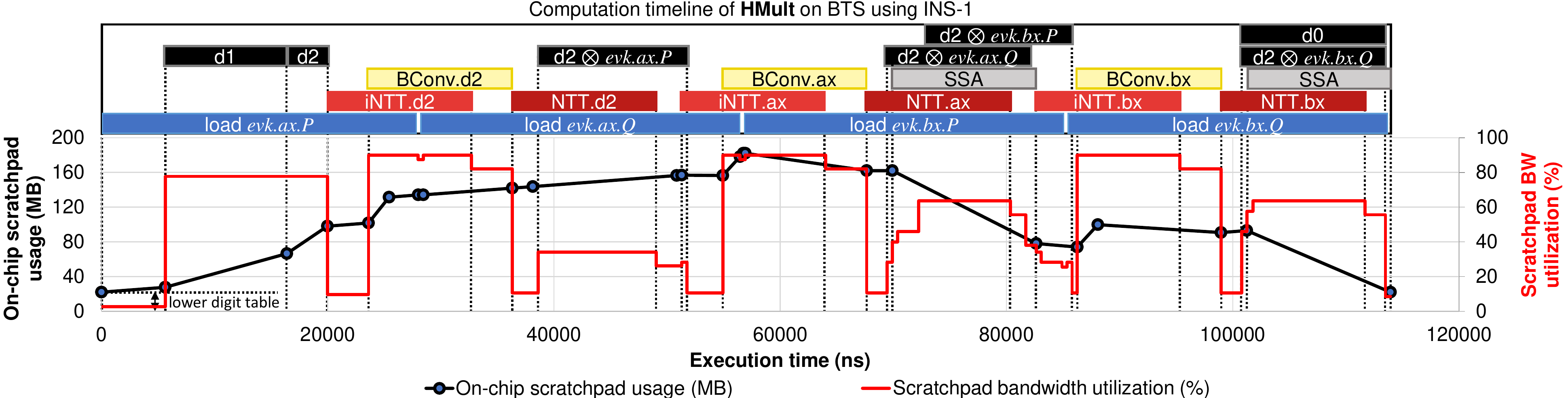}
    \end{center}
    \vspace{-0.1in}
    \caption{Timeline, on-chip scratchpad usage change, and scratchpad bandwidth utilization change when \name performs HMult with \texttt{INS-1}.}
    \vspace{-0.05in}
    \label{fig:gantt-chart}
\end{figure*}

\begin{figure}[tb!] 
    \begin{center}
    \includegraphics[width=0.99\columnwidth]{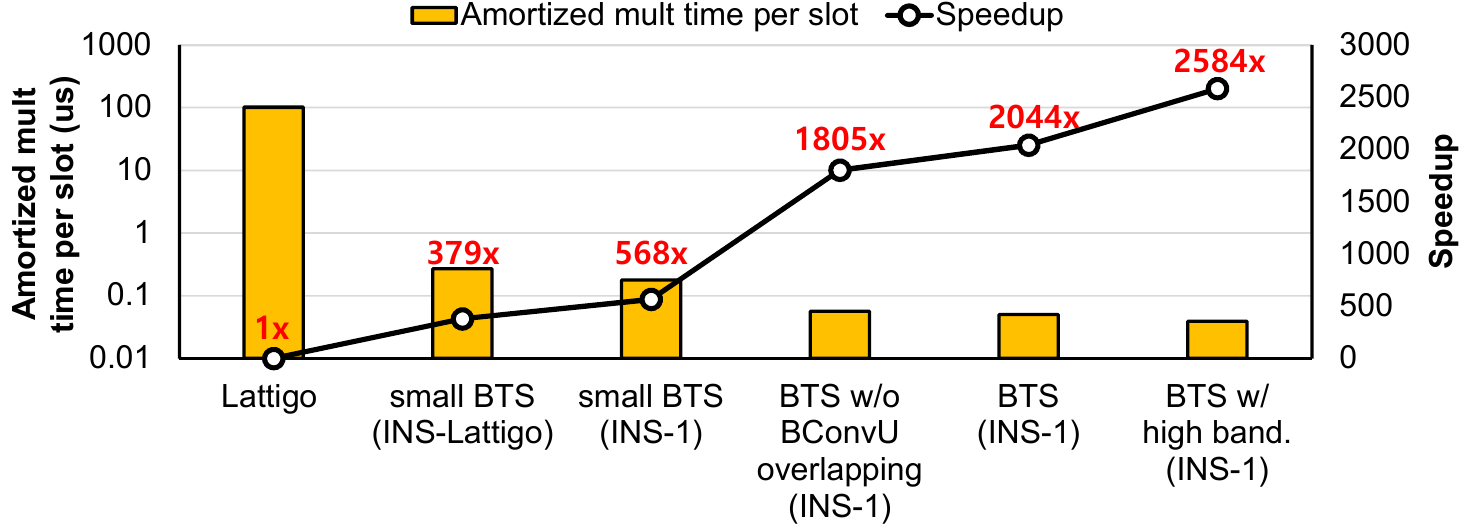}
    \end{center}
    \vspace{-0.1in}
    \caption{The performance and speedup of \amormultslot of BTS when applying various components incrementally. Small BTS is BTS with just enough scratchpad to hold the temporal data of the HE op with no overlapping between BConv and iNTT. The CKKS instance 
    is specified in parentheses.}
    \vspace{-0.19in}
    \label{fig:obligation}
\end{figure}

\noindent
\textbf{Parameter selection in retrospect:}
In Section~\ref{sec:3_contribution1}, we estimated the \amormultslot of CKKS instances assuming an always-hit scratchpad and used it as a proxy for the performance of FHE applications with frequent bootstrapping. While the \amormultslot result from the simulator does not directly match the estimation, the 2GB scratchpad case (Fig~\ref{fig:theoretical}(a)) does concur. This is because the temporal data of \namethree constitutes the largest set (Table~\ref{tab:parameter}) and the corresponding hit rate is affected by the scratchpad capacity.

However, \amormultslot does not always translate to the application performance for the following reasons.
First, when the portion of bootstrapping is relatively small as in ResNet-20 (Fig~\ref{fig:theoretical}(b)), the complexity of HE ops becomes more influential, and a smaller \dnum value is better (\nameone in Table~\ref{tab:resnet-20}).
Second, the better \amormultslot caused by deeper levels from higher \dnums does not translate to better performance when there exists a level imbalance between \cts.
Such an imbalance nullifies the benefit of more available levels
(see Table~\ref{tab:resnet-20} with \nameone and \nametwo).

\noindent
\textbf{PE resource utilization over time:}
Resources populated in PEs are highly utilized while processing HE ops. 
Fig.~\ref{fig:gantt-chart} presents a detailed timeline of \HMult on \nameone when \cts are on the \scratchpad. 
HBM achieves 98\% of its peak bandwidth.
NTTUs are busy processing (i)NTT of three intermediate polynomials (d2, ax, and bx) 76\% of the time. 
BConv is partially pipelined with iNTT and has strong dependency on the subsequent NTT; thus, it occupies BConvU for 33\% of the time.
The scratchpad bandwidth requirement of BConv is high because it must load the partial sum for all $p_i$s in Eq.~\ref{eq:BConv_modified} within $\lsub$ epochs.
BConvU runs SSA while not occupied by BConv.

The bandwidth and capacity utilization of the \scratchpad fluctuate over time while being properly provisioned to meet the requirements.
The average bandwidth usage was 58.6\% over time, peaking at 90\% when processing a BConv.
The required capacity was also highest at BConv.ax at 183MB. 

\begin{figure}[tb!] 
    \begin{center}
    \includegraphics[width=0.99\columnwidth]{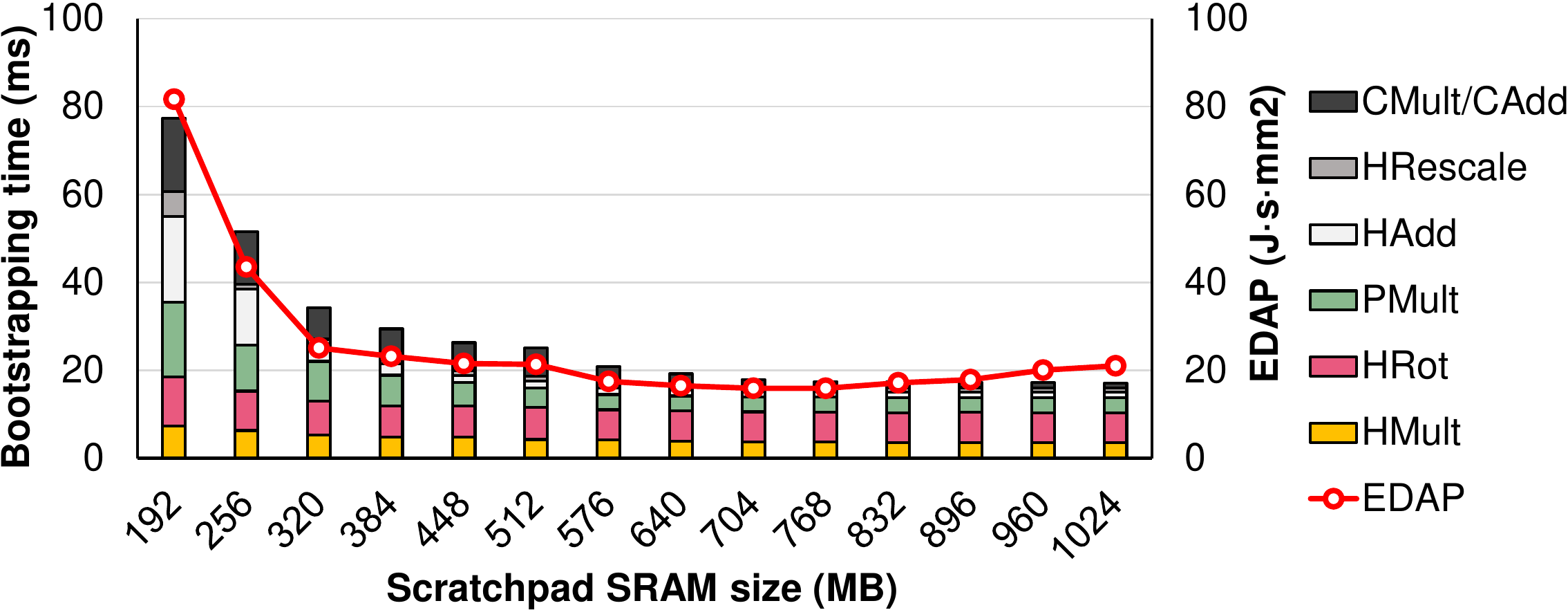}
    \end{center}
    \vspace{-0.1in}
    \caption{The bootstrapping time and Energy-Delay Area Product (EDAP) of \texttt{BTS-1} at various scratchpad SRAM sizes.}
    \vspace{-0.17in}
    \label{fig:EDAP}
\end{figure}

\noindent
\textbf{Ablation study:}
To evaluate the impact of various attributes of \name on its performance,
first we evaluated a small baseline \name ($<$ 230mm\textsuperscript{2}) with just enough scratchpad to hold the temporary data that use \lattigo's CKKS instance ($N=2^{16}$) and without overlapping between BConv and iNTT. The results are 379$\times$ faster \amormultslot compared to \lattigo. 
We incrementally changed the CKKS instance to \nameone and then increased the scratchpad size to 512MB. 
These changes resulted in 1.50$\times$ and 3.18$\times$ speedups, respectively (see Fig.~\ref{fig:obligation}). Finally, additionally overlapping BConv and iNTT results in a 1.13$\times$ speedup, reaching a total of 2044$\times$ speedup compared to \lattigo. 

We also evaluated \name with an HBM bandwidth of 2TB/s.
We reduced the scratchpad size to make room for the added HBM2e PHYs so that \name retains the same total area. The result only shows a 1.26$\times$ speedup as a larger fraction of time is bound to computations, despite the fact that \evk load time is halved.

\noindent
\textbf{Slowdown of FHE:} FHE applications on \name are still slower than their unencrypted counterparts.
HELR is 141$\times$ slower and ResNet-20 inference is 440$\times$ slower compared to when they are run on a CPU system without FHE.
Evaluation of non-polynomial functions such as ReLU, which are costly to evaluate on FHE~\cite{arxiv-2021-relu} results in a greater slowdown for ResNet-20.
Thus, it is crucial to optimize applications to make them more FHE-friendly.

\noindent
\textbf{Impact of the \scratchpad size on the performance and EDAP:}
The performance and energy efficiency of \name improves as we deploy a larger \scratchpad, however becoming saturated as the \scratchpad holds most of the HE ops' working sets.
Fig.~\ref{fig:EDAP} shows the execution time breakdown and energy-delay-area product (EDAP~\cite{isca-2008-cacti}) for the bootstrapping of \nameone with various \scratchpad sizes.
We increased the \scratchpad size 
from 192MB (close to the temporary data for \HMult) by 64MB, up to 1GB.

With a 192MB \scratchpad, \name frequently load \cts from off-chip memory due to capacity misses. At this point, \HMult/\HRot, which used to be dominant (77\% of the bootstrapping time for \lattigo) due to its high computational complexity, now only requires 24\% of the execution time. The rest attributes to \PMult, \HAdd, \HRes, and \CMult/\CAdd. While \name greatly reduces the computation time of \HMult/\HRot with its abundant PEs, the \ct load time, which any HE ops require when SW cache misses occur, is now dominant.

As the \scratchpad size increases, the portion of \HMult/\HRot on bootstrapping increases. This occurs because the SW cache hit rate of \cts for every HE op gradually increases; 65.6\%, 98.8\%, 93.7\%, 98.6\%, 97.5\%, and 47.8\%, for \HMult, \HRot, \PMult, \HAdd, \HRes, and \CMult/\CAdd, respectively, with a 512MB \scratchpad. The execution time of \HMult/\HRot has a lower-bound of the \evk load time, even during SW cache hits.
However, the other HE ops not requiring \evk can take significantly less time due to the ratio of the on-chip over the off-chip bandwidth ($>\!10$), when the necessary \cts are located on the \scratchpad.

%% file: 7_relatedwork.tex
\section{Related Work}
\label{sec:7_related_work}

\noindent \textbf{CPU acceleration}: \cite{github-heaan} parallelized HE ops by multi-threading.
\cite{access-2021-demystify, wahc-2021-hexl} leveraged short-SIMD support. 
\cite{github-lattigo} exploited the algorithmic analysis from \cite{eurocrypt-2021-efficient} for efficient bootstrapping implementation.
Yet other platforms outperform CPU implementations.

\noindent \textbf{GPU acceleration}:
GPUs are a good fit for accelerating HE ops as they are equipped with a massive number of integer units and abundant memory bandwidth.
However, a majority of prior works did not handle bootstrapping~\cite{tetc-2019-bfv,tches-2018-fv,access-2020-privft,access-2021-demystify}.
\cite{tches-2021-100x} was the first work that supported CKKS bootstrapping on GPU.
By fusing GPU kernels, \cite{tches-2021-100x} reduced off-chip accesses and achieved
242$\times$ faster bootstrapping over a CPU.
However, the lack of on-chip storage forces some kernels to remain unfused~\cite{iiswc-2020-ntt}.
\name holds all temporary data on-chip, minimizing off-chip accesses.

\balance
\noindent\textbf{FPGA/ASIC acceleration}:
A different set of works accelerate HE using FPGA or ASIC, but most of them did not consider bootstrapping~\cite{asplos-2020-heax, hpca-2019-roy, fccm-2020-sunwoong-ntt, reconfig-2019-sunwoong-modmult, hpca-2021-cheetah}.
HEAX~\cite{asplos-2020-heax} dedicated hardware for CKKS mult on FPGA, reaching a 200$\times$ performance gain over a CPU implementation.
However, its design is fixed to a limited set of parameters and does not consider bootstrapping.
Cheetah~\cite{hpca-2021-cheetah} introduced algorithmic optimization for an HE-based DNN and proposed an accelerator design suitable for this.
Instead of bootstrapping, Cheetah uses multi-party computation (MPC) to mitigate errors during the HE operation. Cheetah sends a ciphertext with error back to the clients and the clients recrypt it as a fresh ciphertext.
In MPC, the network latency from the frequent communication with the client limits the performance~\cite{van2021practical}, thus introducing a different challenge compared to FHE.
The accelerator design of Cheetah targets a small ciphertext for MPC, which is not suitable for FHE~\cite{micro-2021-f1}.
\fone~\cite{micro-2021-f1} is the first ASIC design that partially supports bootstrapping.
It is a programmable accelerator supporting multiple FHE schemes, including CKKS and BGV.
\fone achieves impressive performance on various LHE applications as it provides tailored high-throughput computation units and stores \evks on-chip, minimizing the number of off-chip accesses.
However, \fone targets the parameter sets with low degree $N$, thus supporting only non-packed (single-slot) bootstrapping, the throughput of which is greatly exacerbated compared to \name.
\fone is 151.4mm\textsuperscript{2} in size at a 12/14nm technology node and shows a TDP of 180.4W excluding the HBM power.

%% file: 8_conclusion.tex
\section{Conclusion}
\label{sec:conclusion}

We have proposed an accelerator architecture for fully homomorphic encryption (FHE),
primarily optimized for the throughput of bootstrapping encrypted data.
By analyzing the impact of selecting key parameter values on the bootstrapping performance of CKKS, an emerging HE scheme, we devised the design principles of bootstrappable HE accelerators and suggested \name, which distributes massively-parallel processing elements connected through a network-on-chip design tailored to the unique traffic patterns of number theoretic transform and automorphism, the critical functions of HE operations.
We designed \name to balance off-chip memory accesses, on-chip data reusability, and the computations required for bootstrapping. 
With \name, we obtained a speedup of 2,237$\times$ in HE multiplication throughput and 5,556$\times$ in CNN inference compared to the state-of-the-art CPU implementations.